\documentclass[aps,prd,twoside,onecolumn,superscriptaddress,floatfix,float,nofootinbib,showpacs]{revtex4}
\usepackage{amsmath}
\usepackage{bm}
\usepackage{xcolor}
\usepackage{hyperref}
\usepackage{url}
\usepackage{graphicx}
\usepackage[large]{subfigure}
\usepackage{amssymb}
\usepackage[amssymb]{SIunits}
\usepackage{aas_macros}
\usepackage{natbib}

\usepackage{multirow}

\newcommand\beqn{\begin{eqnarray}}
\newcommand\eeqn{\end{eqnarray}}

\newcommand\tc{T_{\rm CMB}}
\newcommand\ti{T_{\rm CIB}}
\newcommand\phic{\phi_{\rm CMB}}
\newcommand\phii{\phi_{\rm CIB}}

\def\x{{\bf x}}

\def\k{{\bf k}}
\def\K{{\bf K}}

\def\q{{\bf q}}

\newcommand\lsim{\mathrel{\rlap{\lower4pt\hbox{\hskip1pt$\sim$}}
        \raise1pt\hbox{$<$}}}
\newcommand\gsim{\mathrel{\rlap{\lower4pt\hbox{\hskip1pt$\sim$}}
        \raise1pt\hbox{$>$}}}

\usepackage[mathscr]{eucal}	
\DeclareTextSymbol{\degre}{T1}{23}

\newcommand{\vl} { {\mbf{\ell}} }
\newcommand{\vL} { {\mbf{L}} }

\newcommand{\n} {\hat{\mbf{n}}}

\newcommand{\beq} {\begin{equation}}
\newcommand{\eeq} {\end{equation}}
\newcommand{\bal} {\begin{aligned}}
\newcommand{\eal} {\end{aligned}}

\newcommand{\mbf}[1]{\mbox{\boldmath$#1$}}

\begin{document}

\title{Weak Lensing of Intensity Mapping: the Cosmic Infrared Background}

\author{Emmanuel Schaan}
\email{eschaan@lbl.gov}
\affiliation{Lawrence Berkeley National Laboratory, One Cyclotron Road, Berkeley, CA 94720, USA}
\affiliation{Berkeley Center for Cosmological Physics, University of California, Berkeley, CA 94720, USA}
\affiliation{Department of Astrophysical Sciences, Princeton University, Peyton Hall, Princeton NJ 08544, USA}
\author{Simone Ferraro}
\affiliation{Department of Astronomy and Miller Institute, University of California, Berkeley, CA 94720, USA}
\affiliation{Berkeley Center for Cosmological Physics, University of California, Berkeley, CA 94720, USA}
\author{David N. Spergel}
\affiliation{Department of Astrophysical Sciences, Princeton University, Peyton Hall, Princeton NJ 08544, USA}
\affiliation{Center for Computational Astrophysics, Flatiron Institute, 162 5th Avenue, 10010, New York, NY, USA}

\begin{abstract} 

Gravitational lensing deflects the paths of cosmic infrared background (CIB) photons, leaving a measurable imprint on CIB maps. 
The resulting statistical anisotropy can be used to reconstruct the matter distribution out to the redshifts of CIB sources.
To this end, we generalize the CMB lensing quadratic estimator to any weakly non-Gaussian source field,
by deriving the optimal lensing weights.
 We point out the additional noise and bias caused by the non-Gaussianity and the `self-lensing' of the source field.
We propose methods to reduce, subtract or model
 these non-Gaussianities.
We show that CIB lensing should be detectable with Planck data, and detectable at high significance for future CMB experiments like CCAT-Prime.
The CIB thus constitutes a new source image for lensing studies, providing constraints on the amplitude of structure at intermediate redshifts between galaxies and the CMB.
CIB lensing measurements will also give valuable information on the star formation history in the universe,
constraining CIB halo models beyond the CIB power spectrum.
By laying out a detailed treatment of lens reconstruction from a weakly non-Gaussian source field,
this work constitutes a stepping stone towards lens reconstruction from continuum or line intensity mapping data,
such as the Lyman-alpha emission, absorption, and the 21cm radiation.

\end{abstract}

\maketitle

\section{Introduction}

Weak gravitational lensing probes the projected mass distribution between the source and the observer, and is therefore sensitive to the underlying cosmology. 
Using multiple source and lens redshift bins further allows us to track the amplitude of structure across cosmic time.  
Lensing introduces subtle correlations that would be forbidden by the assumption of statistical isotropy and homogeneity, and these correlations can be used to reconstruct mass maps, which include the combined effect of dark matter, baryons, neutrinos and all other forms of energy density.  
At the same time, the amplitude of these fluctuations and their angular size on the sky are determined by the expansion history and the nature of the gravitational force, making weak lensing also a sensitive probe of dark energy, modified gravity and the masses of neutrinos.  It is therefore one of the most promising cosmological tools for the decades to come.

So far, two phenomenologically distinct regimes of weak lensing have been explored: on the one hand, optical surveys use images of individually resolved galaxies to measure small correlations in their ellipticities induced by lensing (see \cite{2001PhR...340..291B, 2005astro.ph..9252S, 2015RPPh...78h6901K} for a review).  In this case, the source plane is highly non-Gaussian\footnote{Meaning that the pixel-to-pixel joint probability distribution is highly non-Gaussian.}, and measurements of shear of individual galaxies is appropriate.  Galaxy lensing has been detected in a large number of surveys, including most recently by the Kilo Degree Survey (KiDS, \cite{2017MNRAS.465.1454H}), the Dark Energy Survey (DES, \cite{2016PhRvD..94b2002B}) and the Hyper Supreme-Cam(HSC, \cite{2017arXiv170506745M}).

In the opposite regime, when the source is a Gaussian random field such as the Cosmic Microwave Background (CMB) radiation, 
a rich theory of estimators has been developed (\cite{2006PhR...429....1L, 2010GReGr..42.2197H} for a review), the most commonly used being the quadratic estimator (QE) of Hu and Okamoto \cite{2002ApJ...574..566H}.  
While only optimal among the class of estimators that are quadratic in the measured lensed fluctuations\footnote{More general, but much more computationally expensive methods can be used and they have been shown to perform better than the quadratic estimator in the low-noise or small-scale regimes \cite{2003PhRvD..68h3002H, 2017arXiv170408230C}.}, it has been shown to be close to optimal for the analysis of current experiments.  Recent detections include those by the Atacama Cosmology Telescope (ACTPol, \cite{2016arXiv161109753S}), the South Pole Telescope (SPT-SZ, \cite{2017arXiv170500743O}), POLARBEAR \cite{2014PhRvL.113b1301A} and the Planck mission \cite{2016A&A...594A..15P}.
In the case of CMB lensing, the statistical properties of the source field are very well characterized, and so is its redshift. 
It is possible to extract information from both the shearing of the fluctuations (which are on average round for the unlensed CMB) and a local magnification or demagnification.  
In fact, on large scales, the quadratic estimator can be rewritten as a minimum variance combination of estimators of shear and dilation \cite{2012PhRvD..85d3016B}.

The cosmic infrared background (CIB) \cite{1996A&A...308L...5P} is the highly blended and unresolved thermal emission from a large population of dusty star-forming galaxies over a broad range of redshifts $1 \lesssim z \lesssim 4$. The CIB is an excellent probe of star formation history, and its fluctuations reflect the clustering properties of the underlying galaxy populations.
Indeed, the CIB fluctuations are highly correlated with the CMB lensing potential, meaning that it closely traces the dark matter distribution at those intermediate redshifts \cite{2014A&A...571A..18P}.
Its source redshift distribution is broad, and current models predict that observations at lower frequency receive a larger contribution from higher redshifts \cite{2014MNRAS.439..123L, 2014MNRAS.439..143P}.

A simple geometric argument shows that the sensitivity to a fixed mass lens is maximized when the lens is roughly half-way between the source and the observer.  
Having source images at different redshifts therefore allows to reconstruct the mass distribution in the universe tomographically.
The CIB source distribution is at intermediate redshifts between the galaxies typically used in galaxy lensing and the CMB, thus providing a useful complement.

Because of the central limit theorem, CIB fluctuations are closer to Gaussian than images of resolved galaxies, but are weakly non-Gaussian nonetheless \cite{2014A&A...571A..30P}. 
In this paper, we generalize the CMB lensing quadratic estimator to the case of any weakly non-Gaussian source field, and apply this formalism to the CIB.
We  discuss how the non-Gaussianity of the source modifies the usual noise biases and how new biases arise due to the large redshift span
of the non-Gaussian emission. 
Specifically, galaxies at low redshift that source the CIB emission also act as lenses for the emission from higher redshift CIB galaxies, an effect that we call `self-lensing'.  This affects both the power spectrum of the inferred CIB lensing potential and its cross-correlation with low-redshift tracers.
We explore methods to mitigate this self-lensing bias.

While we take the CIB as our primary example, our formalism is general and applies to lensing of any other weakly non-Gaussian sources, such as the Lyman-alpha forest \cite{2017arXiv170607870C, 2017arXiv170608939M}, the 21cm radiation from either the epoch of reionization or low-redshift galaxies \cite{2006ApJ...653..922Z, 2015MNRAS.448.2368P, 2008MNRAS.388.1819L, 2010PhRvD..81l3015L}, or any other line intensity mapping.

Finally, any residual CIB contamination in foreground cleaned CMB maps is known to bias CMB lensing reconstruction \cite{2014ApJ...786...13V, 2014JCAP...03..024O}. 
However, the fact that the CIB field is itself lensed introduces an additional bias, not considered previously, which we point out and discuss in this paper.

This paper is organized as follows. 
We begin with a heuristic review of lensing estimation in Sec.~\ref{sec:heuristic}, in order to motivate our choice of a quadratic estimator for lensing.
In Sec.~\ref{sec:formalism}, we describe our model for the auto and cross-power spectra for various observables, including galaxies, the lensing convergence, and the CIB. 
In Sec.~\ref{sec:ps}, we compute the lensing correction to the CIB power spectrum. 
In Sec.~\ref{sec:QE} we review the quadratic estimator formalism and extend it to non-Gaussian source fields.
In Sec.~\ref{sec:forecasts} we present forecasts for current and future experiments, showing that the signal should be detectable at high significance. 
In Sec.~\ref{sec:biases} we explore how the usual lensing biases are modified in the case of a non-Gaussian source, and show that new biases are generated by the extended and non-Gaussian nature of the CIB.  
Finally, in Sec. \ref{sec:biasCMB}, we discuss the bias to CMB lensing given some residual level of CIB contamination to CMB maps, taking into account that the CIB itself is lensed. 
We conclude in Sec.~\ref{sec:conclusion}.
Appendices~\ref{app:cib_halo_model}-\ref{app:trispectrum_measurement} present the details of the CIB halo model implemented in this paper, a full derivation of the  lensing kernel for CIB lensing, a derivation of the lensing signal-to-noise in a simple limiting (Poissonian) regime, and the method used for our direct measurement of the CIB trispectrum on the Planck GNILC maps.

Throughout the paper, we will assume a flat $\Lambda$CDM cosmology with cosmological parameters from the 2015 Planck release (column 3 of Table 4 of
 \cite{2016A&A...594A..13P}).

\section{Heuristic review of lensing estimation}
\label{sec:heuristic}

\begin{figure}[h]
\centering
\includegraphics[width=0.9\columnwidth]{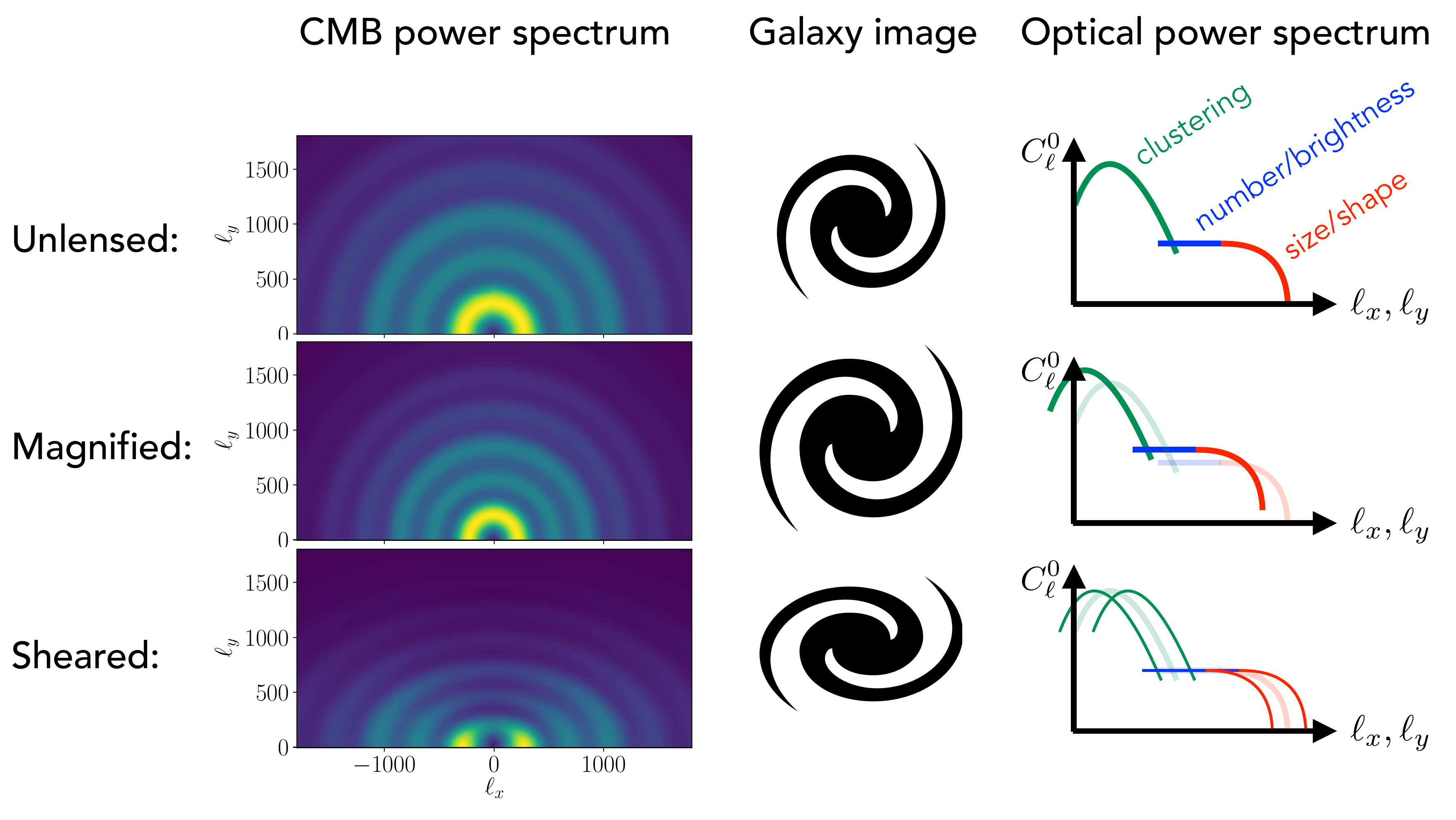}
\caption{
Schematic illustration of the `large-scale lens' regime, in which the unlensed image varies on much smaller scales than the lensing convergence field. 
In this regime, shear and convergence are uniform on the scale of several galaxies, or several CMB hot or cold spots.
The figure shows the analogy between galaxy lensing and CMB lensing estimators in this regime. 
\textbf{Left column:} The large-scale lens regime is one of the regimes in which the CMB lensing quadratic estimator operates. In this regime, the quadratic estimator can be shown to look for monopolar (magnification) and quadrupolar (shear) distortions in the local observed power spectrum \cite{2012PhRvD..85d3016B, 2017arXiv170902227P}. 
\textbf{Central column:} 
The shear is estimated from the galaxy shape (quadrupole of the image),
and in principle magnification from the galaxy size, brightness or number density (monopole of the image). 
\textbf{Right column:} Na\"ive schematic of the power spectrum of an optical image, on a field containing galaxies and with uniform magnification/shear. 
We schematically describe the power spectrum as a clustering component, plus a 1-halo or 1-galaxy term. 
The amplitude of this 1-halo/1-galaxy encodes both the galaxy number density and brightness, and its turnover encodes the galaxy size and shape.
The effect of magnification is to rescale the multipoles $\ell_x, \ell_y$ isotropically, as well as the power spectrum amplitude. 
Magnification thus affects the local number density, brightness and size of the galaxies, without distorting their shapes. 
On the other hand, the effect of shear is an anisotropic rescaling of the multipoles $\ell_x$ and $\ell_y$, leaving the number density and brightness unchanged. 
This figure shows that the shape, size, brightness and number density measured from individual objects (individual galaxies or CMB hot spots) is also encoded in the power spectrum of these objects (power spectrum of the galaxy field image or the CMB).
In the large-scale lens regime, the information measured by galaxy lensing estimators on individual objects is completely analogous to that measured by the quadratic estimator on the CMB power spectrum.
}
\label{fig:cartoon_analogy_lensing}
\end{figure}

In this section, we present a heuristic review of the lensing estimation methods used for the CMB and galaxies.
Understanding the lensing estimators in the two limiting regimes, from discrete point-like objects to a continuum Gaussian random field, will inform us on what estimator to use in the intermediate case of the CIB.
The goal of this section is therefore to provide motivation for the CIB CIB quadratic lensing estimator we present below.
Lens reconstruction consists of inferring the unlensed map and the convergence map, given the observed lensed map.
In the case of CMB lensing, this problem is well posed because the statistics of the unlensed CMB is largely understood: it is a Gaussian random field with a known power spectrum.
Thanks to this prior on the unlensed map, the exact posterior for the lensing map can be written explicitly \cite{2003PhRvD..68h3002H}, and numerical exploration of this posterior for realistic datasets is possible \cite{2017arXiv170408230C, 2017arXiv170806753M}. 

On the contrary, in the case of galaxy lensing, the unlensed images are non-Gaussian and complex, and a full prior on the unlensed field is not readily available.
One way around this is to assume a reasonable partial prior.
For instance, barring intrinsic alignments, unlensed galaxy ellipticities are assumed to be uncorrelated.
One builds a catalog of the individual observed galaxy ellipticities, and infers the shear.
Another partial prior is that galaxy sizes and brightness are uncorrelated on large scales. 
Yet another one is that galaxy positions at high and low redshift should be uncorrelated.
These priors can be used to detect magnification \cite{2005ApJ...633..589S, 2011arXiv1111.1070H, 2011MNRAS.415.3485Y, 2012ApJ...744L..22S}.

Consider now the `large-scale lens regime', where the lensing field varies on scales larger than the typical fluctuations in the unlensed image.
This is the regime of galaxy lensing, where shear and convergence are coherent on the scale of several galaxies.
This regime also occurs in CMB lensing, for lensing modes that are coherent over many CMB hot and cold spots.
In this case, the CMB lensing quadratic estimator effectively looks for distortions of the locally measured power spectrum. 
Indeed, for a small patch with roughly uniform shear $\gamma$ and convergence $\kappa$, the local power spectrum is modified as \cite{2012PhRvD..85d3016B}:
\beq
C_\vl = C^0_\ell 
\left[
1
+ \kappa \frac{\partial \ln \ell^2 C^0_\ell}{\partial \ln \ell}
+ \gamma \cos(2\theta_{\ell}) \frac{\partial \ln C^0_\ell}{\partial \ln \ell} 
\right],
\eeq
where $C^0_\ell$ and $C_\ell$ are the unlensed and lensed power spectra respectively, $\kappa$ is the convergence and $\gamma$ the shear amplitude, assumed to be uniform on the patch where the power spectrum is measured. The angle $\theta_\ell$ is the angle between the direction of the shear and the wave vector $\vl$.
As this equation shows, magnification results in a monopole distortion of the 2D power spectrum, and shear produces a quadrupolar distortion.

In this large-scale lens regime, where the CMB quadratic estimator measures shear and magnification, it is close to optimal. 
However, this estimator would be suboptimal
 in galaxy lensing, for several reasons. 
For example, applying a quadratic estimator to the intensity map
of a highly populated galaxy field would implicitly weight galaxies by their brightness, instead of the uncertainty on their shapes. The estimator would thus be dominated by the few brightest galaxies in the field.

The CIB is a somewhat intermediate case.
Similarly to galaxy images, the unlensed CIB is a non-Gaussian field for which a full statistical description is not available.
Without an explicit prior on the unlensed CIB, it is not possible to write down an explicit posterior and explore it, as is done for CMB lensing.
However, the CIB is similar to the CMB in that it is a continuous field, with small fluctuations around a mean value (of order percent).
Furthermore, we do have some knowledge about the non-Gaussianity of the CIB: its bispectrum has been measured \cite{2014A&A...571A..18P}, 
and halo models predicting its trispectrum exist \cite{2014MNRAS.439..123L, 2014MNRAS.439..143P}.
The non-Gaussianity of the CIB is weak, in a sense that we shall define precisely later.
Intuitively, this is expected from the central limit theorem, and the fact that the CIB is the superposition of many blended galaxy emissions.
It is thus natural to build upon a quadratic estimator designed for Gaussian random fields
and derive the optimal weights for any weakly non-Gaussian field. 
We further quantify the statistical error and biases of this estimator, due to the non-Gaussianity of the CIB.
On small scales ($\ell \gtrsim 1000$), the CIB is dominated by the galaxy shot noise, which is expected to be nearly Poissonian.
On these scales, similar to the galaxy lensing case, the  CIB multipoles become highly correlated. 
Accounting for these correlations is then crucial to avoid overcounting the lensing information.

The rest of the paper formalizes this heuristic intuition.
To start, we first need to introduce our modeling of the galaxy number density field, the lensing convergence field, and the CIB temperature fluctuations.
This is done in the next section.

\section{Modeling tracers, lensing potential, and the CIB}
\label{sec:formalism}

\subsection{Galaxy, CIB, lensing fields and their power spectra}

In this section, we define our notations for the auto and cross power spectra of tracers, lensing potential or convergence, and the CIB.
For the number counts of galaxies (or any other tracer of the matter distribution), we denote the fractional fluctuations in number density $n_g$ in direction $\n$ on the sky as $\delta_g(\n) = n_g(\n) / {\bar n}_g - 1$.
Lensing is expressed interchangeably in terms of the lensing potential $\phi$ or convergence $\kappa$,
related by\footnote{Throughout this paper, we adopt the optical lensing sign convention (e.g., \cite{2001PhR...340..291B}) rather than the CMB lensing convention (e.g., \cite{2006PhR...429....1L}). This only affects the sign of the displacement vector and of the lensing potential, but not the sign of the convergence.}
$\kappa =  \frac{1}{2} \nabla^2 \phi$.
We denote by $\kappa_{\rm gal}$, $ \kappa_\text{CMB}$ and $\kappa_\text{CIB}$ the lensing convergence reconstructed from galaxy shapes, the CMB and the CIB respectively.

Each observable $A \in \left\{ \delta_g, \kappa_\text{gal}, \kappa_\text{CIB}, \kappa_\text{CMB} \right\}$ is a projection of the total matter density contrast $\delta_m$ between the source and the observer, weighted by an efficiency kernel $W_A$:
\beq
\label{eq:projection}
A(\mbf{n}) =  \int d\chi \; W^A(\chi) \; \delta_m(\chi \n, \chi),
\eeq
where $\chi$ denotes the comoving radial distance or the comoving angular diameter distance -- identical for a flat universe.

For the projected galaxy overdensity field $\delta_g(\n)$, the efficiency kernel is 
\beq
W^{g}(\chi) = b_g(z) \frac{1}{n_g} \frac{dn_g}{dz} \; \frac{dz}{d\chi},   \text{\ \ \  with  }
n_g = \int dz \frac{dn_g}{dz},
\eeq
where $b_g(z)$ is the galaxy bias and $dn_g/dz$ is the redshift distribution of the galaxies.

In the case of lensing, the integral in Eq. \eqref{eq:projection} is taken along the unperturbed (straight) path between the source and the observer, rather than on the true (perturbed) path.
 This Born approximation has been extensively studied and shown to be an excellent approximation in the regime considered \cite{2016JCAP...08..047P, 2017arXiv170203317F, 2017PhRvD..95l3503P}. 
Given a source at comoving distance $\chi_S$, the lensing efficiency is 
\beq
W^\kappa (\chi, \chi_S) = \frac{3}{2} \left( \frac{H_0}{c} \right)^2 \Omega_m^0 \frac{\chi}{a(\chi)} \left( 1 - \frac{\chi}{\chi_S} \right).
\eeq

The CMB originates from the surface of last scattering at comoving distance $\chi_\text{SLS}$, corresponding to $z_\text{SLS} \sim 1100$.
As a result, the CMB lensing kernel is given by:
\beq
W^{\kappa_\text{CMB}} (\chi) = W^\kappa (\chi, \chi_\text{SLS}).
\eeq

For the lensing of galaxies, the sources are distributed in redshift and the efficiency kernel is obtained by averaging over the source distribution $dn_S/d\chi_S$:
\beq
W^{\kappa_\text{gal}} (\chi) = \;\frac{1}{n_\text{S}} \int d\chi_S \ \frac{dn_\text{S}}{d\chi_S} \; W^\kappa(\chi, \chi_S), \text{\ \ \  with  }
n_S = \int d\chi \frac{dn_S}{d\chi},
\eeq

\begin{figure}[h]
\centering
\includegraphics[width=0.49\columnwidth]{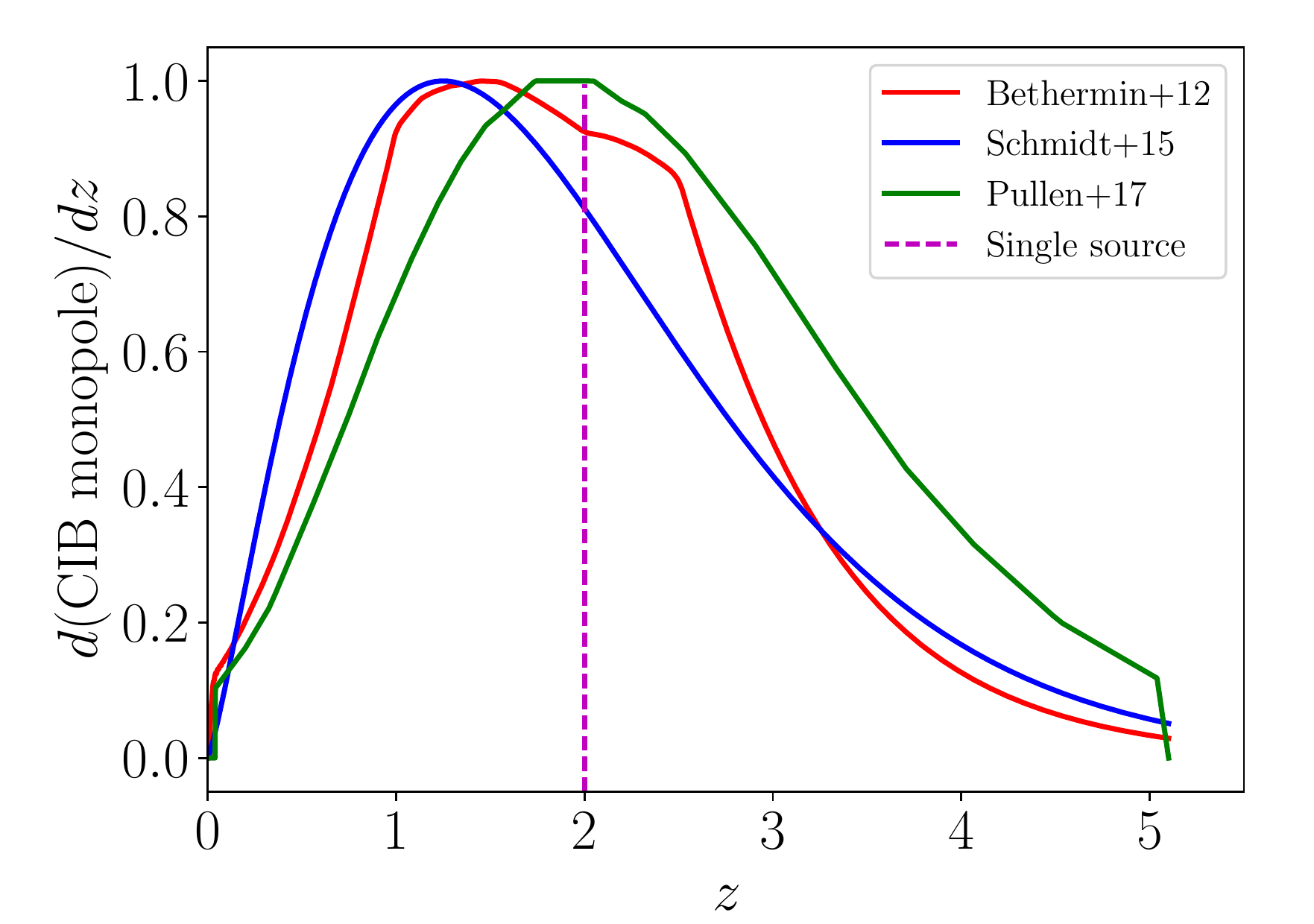}
\includegraphics[width=0.49\columnwidth]{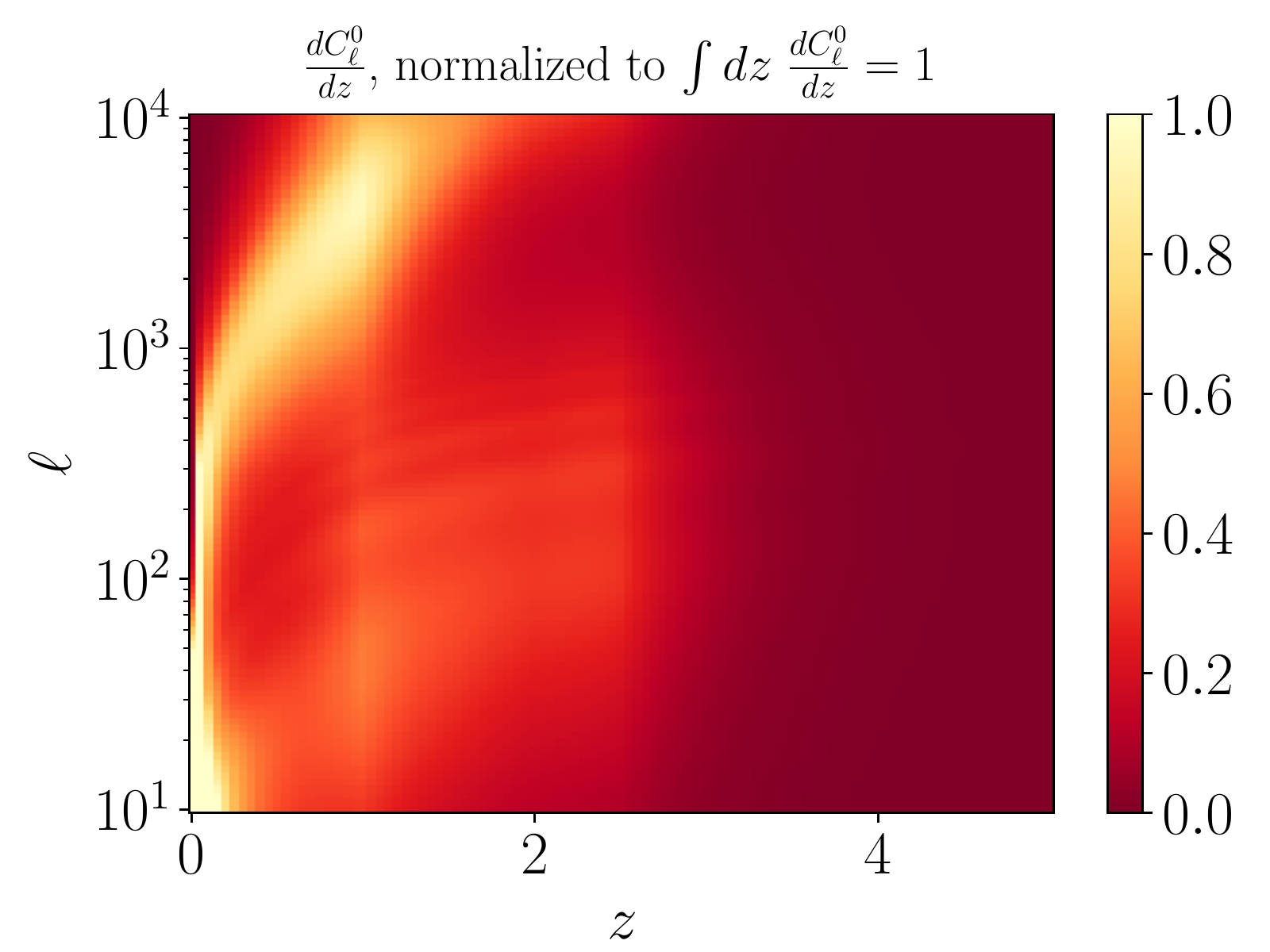}
\caption{
Redshift contribution to the CIB monopole (left panel) and to the CIB fluctuations (right panel) seen by Planck at 545GHz.\\
\textbf{Left:} Redshift dependence of the CIB monopole at $545$ GHz from different halo models:
Bethermin+12 \cite{2012ApJ...757L..23B} relies on the galaxy SEDs measured by Herschel, and reproduces Herschel galaxy counts;
Schmidt+15 \cite{2015MNRAS.446.2696S} is fit to the cross-correlation of Planck HFI data with SDSS quasars.
Pullen+17 \cite{2017arXiv170706172P} is inferred from a fit to the cross-correlation of Planck HFI data with SDSS quasars and CMASS.
This highlights a significant modeling uncertainty on the redshift distribution. \\
\textbf{Right: }Since lensing is reconstructed from the CIB fluctuations rather than the monopole, we plot the redshift dependence of the CIB power spectrum, assuming the halo model of Penin+14 \cite{2014MNRAS.439..143P}, based on the Bethermin+12 model (see left panel). 
Various CIB models will differ to a similar extent as for the CIB monopole (left panel).
Comparing left and right panels, low redshifts ($z\lesssim 0.5$) make a small contribution to the CIB monopole, but a large contribution to the CIB power spectrum. However, these low redshift CIB sources are presumably known galaxies, and could therefore be masked if needed.}
\label{fig:cib_source_dist}
\end{figure}

Similarly, for the CIB, we loosely define a source distribution $W^{\rm CIB}$ and a resulting CIB lensing kernel $W^{\kappa_\text{CIB}}$. 
The case of CIB lensing is slightly more subtle than those of CMB and galaxy lensing. 
Indeed, one might be tempted to use the redshift distribution of the CIB monopole as the relevant CIB source distribution. 
However, as we explain later, 
we are not reconstructing the lensing from the CIB monopole, but instead from the CIB fluctuations.
As a result, the relevant CIB `source distribution' is related to the redshift distribution of the CIB power spectrum, not the monopole (see Fig.~\ref{fig:cib_source_dist}).
The exact expression for the appropriate CIB source distribution is derived and discussed in App. \ref{app:cib_lensing_kernel}. 
Due to the very extended nature of the CIB emission, there is a non-negligible overlap between the CIB emission kernel and the CIB lensing efficiency, causing `self-lensing' biases that we discuss later.
Here, approximating the CIB as a single source at redshift 2 is sufficient for estimating the lensing signal. 
However, we will take into account the large width of the CIB redshift distribution when assessing biases in the lens reconstruction.

In the Limber and flat sky approximations,
the cross-spectrum $C_\ell^{AB}$ between observables $A$ and $B$ is related to the matter power spectrum $P_m$ via
\beq
C_\ell^{AB} = \int \frac{d\chi}{\chi^2} \; W^A(\chi)W^B(\chi) \; P_m\left(k = \frac{\ell +1/2}{\chi}, \chi \right).
\eeq

\begin{figure}[h]
\centering
\includegraphics[width=0.49\columnwidth]{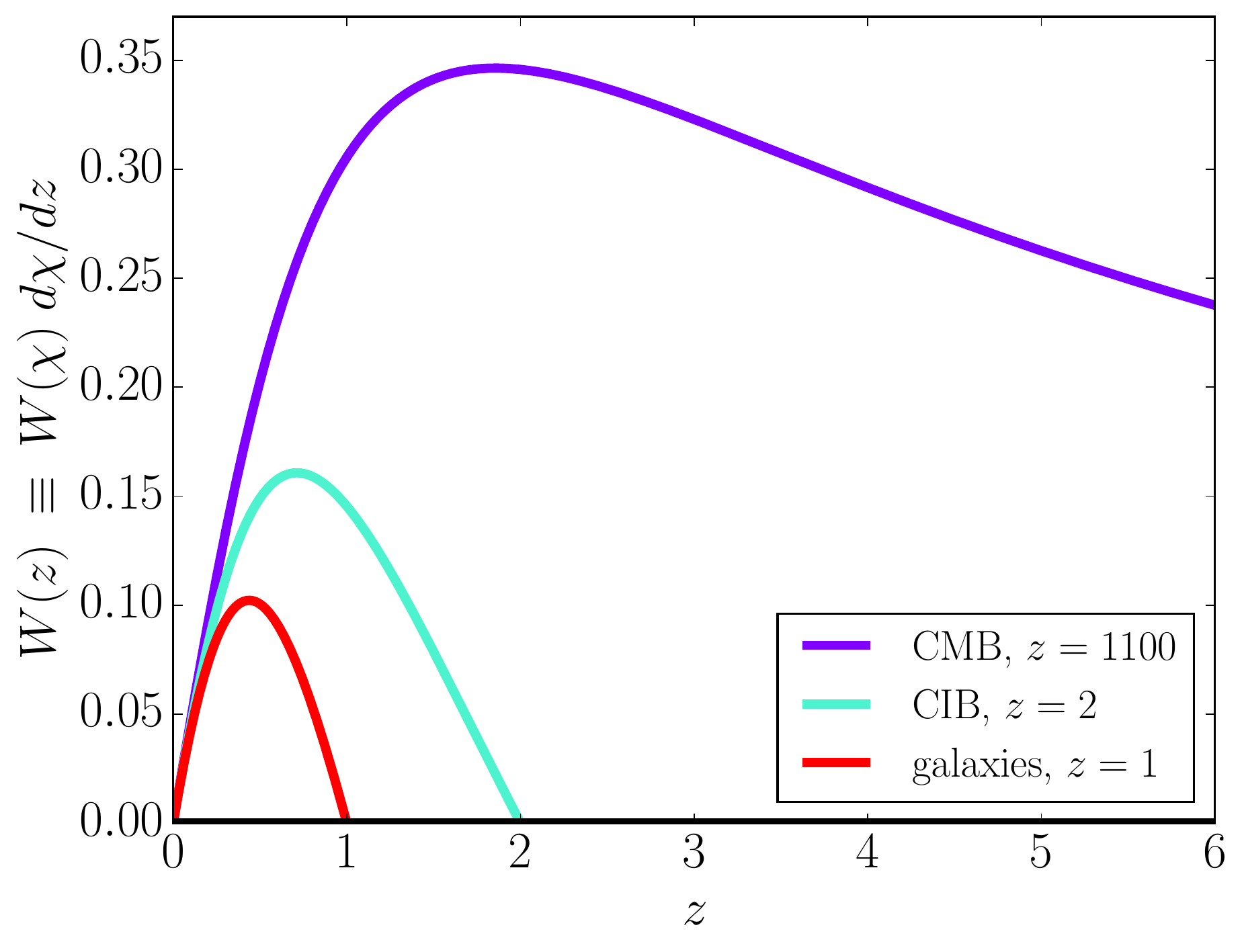}
\includegraphics[width=0.49\columnwidth]{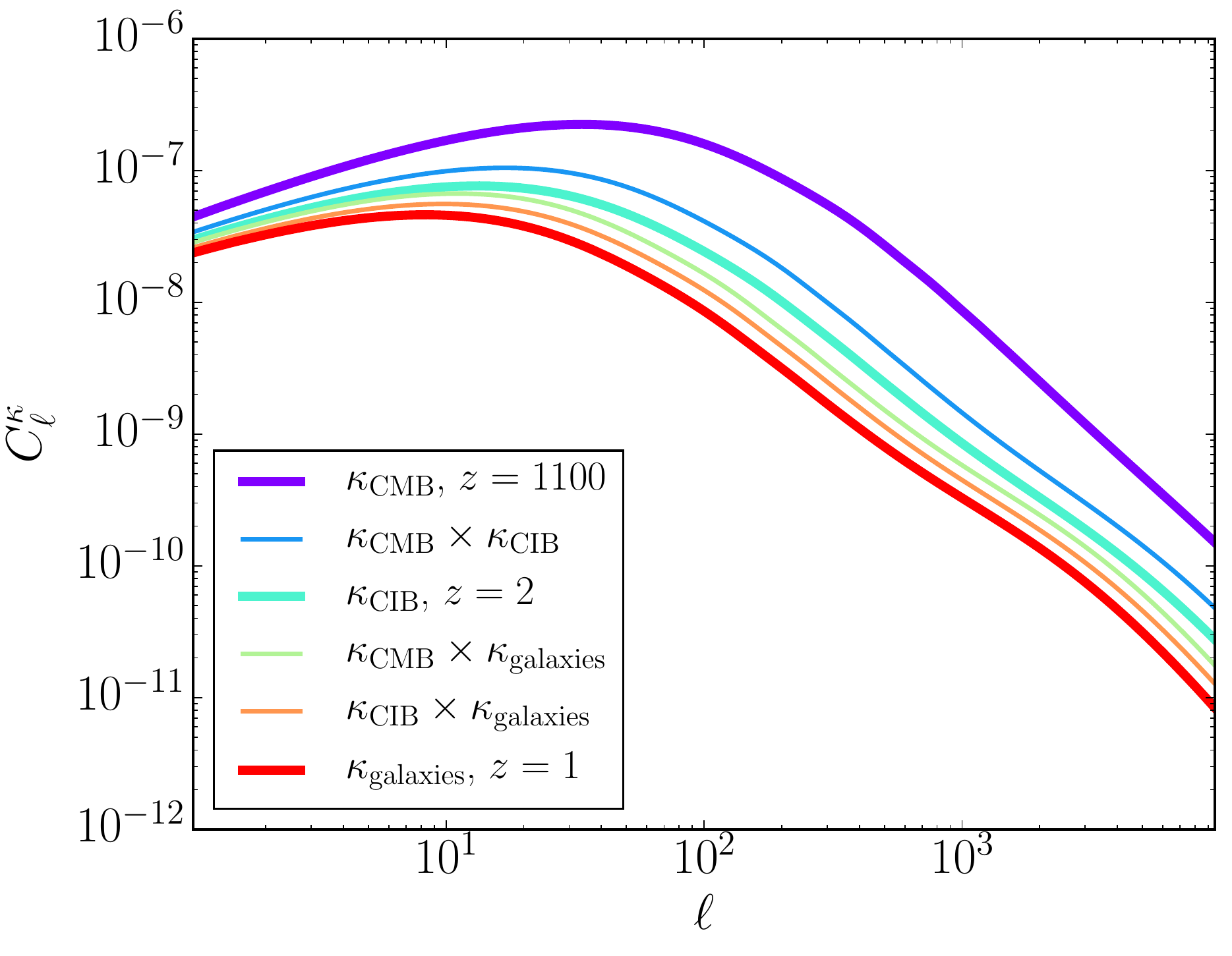}
\caption{
\textbf{Left:} lensing efficiency kernels for the lensing of the CMB, CIB (assuming a single $z_S=2$ source plane) and galaxies (assuming a single $z_S=1$ source plane). 
CIB lensing probes an intermediate redshift range between CMB and galaxy lensing.\\
\textbf{Right:} Auto- and cross-power spectra of the lensing convergence for galaxies, the CMB and the CIB. 
For lower redshift sources, the lensing power is smaller, peaks on larger scales, and has a more visible `1-halo' contribution at $\ell\gtrsim1000$.
}
\label{fig:lensing_kernels}
\end{figure}

We show the lensing auto and cross-power spectra in Fig.~\ref{fig:lensing_kernels}. For lower redshift sources, the lensing power is smaller, peaks on larger angular scales, and has a more visible `1-halo' contribution at $\ell>1000$. For reference, the root-mean square lensing deflection is $1.3'$ for CIB at $z=2$, compared to $2.3'$ for CMB at $z=1100$. 

\subsection{Cosmic infrared background: data and halo model}

Several halo model prescriptions exist for the CIB \cite{2012A&A...537A.137P, 2014MNRAS.439..143P, 2011A&A...529A...4B, 2012ApJ...757L..23B, 2013MNRAS.436.1896A, 2017MNRAS.466.4651W, 2017arXiv170706172P, 2015MNRAS.446.2696S, 2018MNRAS.tmp...57W}.
They differ in terms of their assumptions, their level of realism and complexity.
While most successfully reproduce the observed CIB power spectrum, their best fit parameters differ in detail (see, for instance, the redshift distributions from several models in Fig.~\ref{fig:cib_source_dist}).

We implement the halo model of \cite{2012A&A...537A.137P, 2014MNRAS.439..143P}, summarized in App.~\ref{app:cib_halo_model}, and use it throughout this paper. 
This model includes a halo occupation distribution, which allows to self-consistently predict the 2-halo, 1-halo and galaxy shot noise terms. This also allows to compute all the halo model terms for the higher point functions, such as the CIB trispectrum, most relevant in this paper.
In this model, the average CIB intensity is determined by the galaxy flux distribution function from \cite{2012ApJ...757L..23B}. 
These galaxies are assumed to populate dark matter halos according to the halo occupation distribution (HOD) of \cite{2010ApJ...719...88T}.
The CIB fluctuations are assumed to linearly trace the fluctuations in total galaxy number density,
neglecting the fact that galaxy clustering is luminosity-dependent.

A generic feature of the CIB halo models is that the CIB emissions at different frequencies are produced by galaxies at slightly different redshifts. 
Furthermore, a more sensitive instrument will resolve and typically mask more individual sources. As a result, the observed CIB depends on frequency band and on the flux cut for the instrument. 
In what follows, we consider the CIB as seen by Planck \cite{2016A&A...594A...1P} and by a CCAT-Prime-like experiment\footnote{http://www.ccatobservatory.org/} with a configuration from Table 2 of \cite{2017arXiv170806365M}. The assumed specifications are presented in Tab.~\ref{tab:specs_planck_ccat}.
\begin{table}[h!]
\begin{center}
\begin{tabular}{ |l|c|c|c|c|c|c| } 
\hline
& Frequency & Beam FWHM & White noise & Flux cut & Maximum multipole & $f_\text{sky}$ \\
\hline
Planck & 545GHz & $4.8^\prime$ & 13.5 Jy/rad, i.e. $822 \mu K_\text{CMB}^\prime$ & 350 mJy & 3000 & 0.4\\
CCAT-Prime & 545GHz & $0.5^\prime$ & 1.2 Jy/rad, i.e. $20 \mu K_\text{CMB}^\prime$ & 73 mJy & 40000 & 0.4\\
\hline
\end{tabular}
\end{center}
\caption{Specifications used for Planck \cite{2016A&A...594A...1P} and CCAT-Prime (see Table 2 of \cite{2017arXiv170806365M}). 
The Planck flux cut at 545GHz is taken from Table 2 in \cite{2016A&A...594A...1P}.
For CCAT-Prime, we replaced the frequency 405GHz by 545GHz, in order to use the same halo model for the signal. 
Since the CIB emission is larger at higher frequencies, this will alter the signal-to-noise in the CCAT-Prime band.
We neglect this, given
the large uncertainty on the actual noise level of the cleaned CIB map from CCAT-Prime.
The flux cut for CCAT-Prime is obtained by assuming that point sources detected at $5\sigma$ and above via matched filter are masked.
}
\label{tab:specs_planck_ccat}
\end{table}
For the Planck 545GHz CIB map, we use the 350 mJy flux cut presented in Table 2 of \cite{2016A&A...594A...1P}.
For CCAT-Prime, we estimate the point source detection noise with a matched filter (matched to the CCAT-Prime beam), and assume that the point sources detected at $5\sigma$ have been masked. This corresponds to a flux cut at 73 mJy. 

\begin{figure}[h]
\centering
\includegraphics[width=0.49\columnwidth]{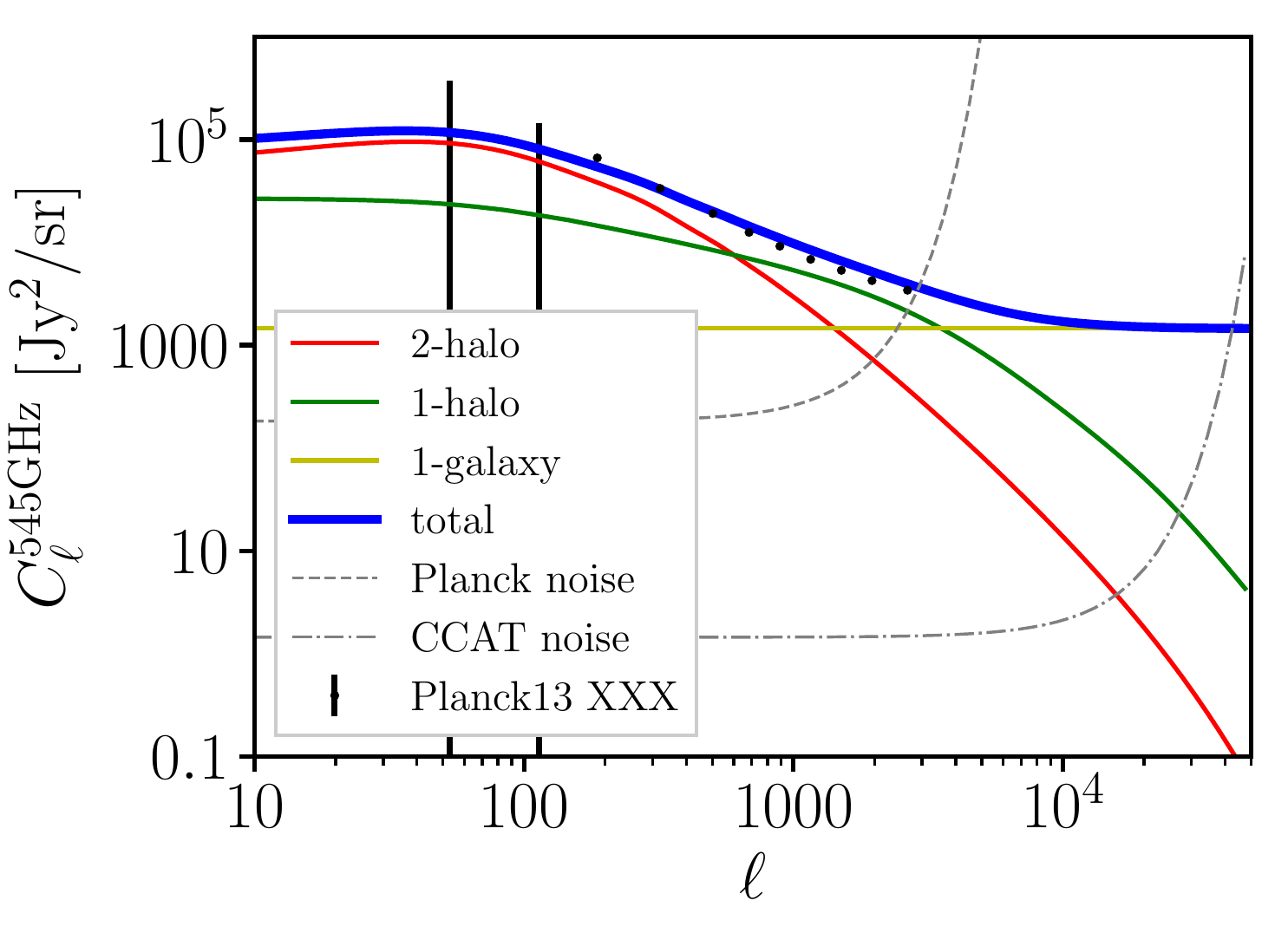}
\includegraphics[width=0.49\columnwidth]{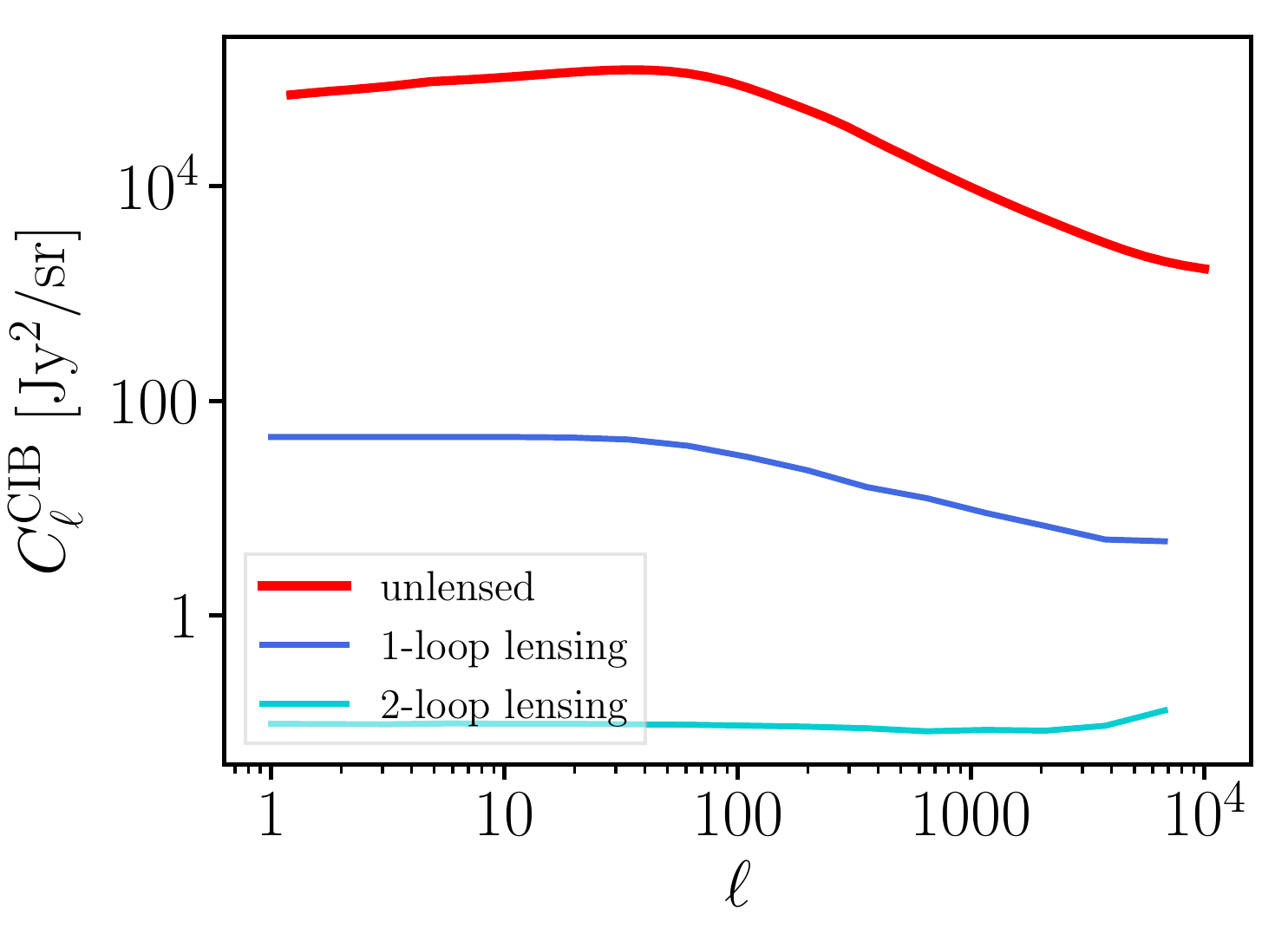}
\caption{
\textbf{Left:} Power spectrum of the CIB at 545GHz as measured by Planck (black points with error bars) and predicted by the halo model of Penin+14 (thick red line).
The halo model calculation is the sum of the 2-halo term (dark orange line), the 1-halo term (light orange line) and the shot noise or 1-galaxy term (yellow line). 
Because they only start to dominate on small scales, the 1-halo and shot noise terms are not very well constrained from the Planck power spectrum alone.
The grey dashed curve shows the noise level for Planck.
The grey dot-dashed curve shows the noise level for CCAT-Prime.
Planck's beam and sensitivity make it a perfect CIB experiment below $\ell \lsim 3000$, while CCAT-Prime will be a perfect CIB experiment out to $\ell \sim 50000$.
These considerations neglect the issue of component separation, and in particular the ability to distinguish the CIB from Galactic dust.\\
\textbf{Right:} Comparison between the unlensed CIB power spectrum (red curve) and the corrections due to weak lensing of the CIB, from Eq.~\eqref{eq:lensed_power}.
The 1-loop term (blue curve) corresponds to the lensing correction at first order in $C_L^{\phi\phi}$, and the 2-loop term (cyan curve) to the lensing correction terms involving two powers of $C_L^{\phi\phi}$.
Weak lensing thus has almost no effect on the CIB power spectrum;
however, this does not mean that lensing cannot be detected, as we show below.
}
\label{fig:cl_cib_545_planck_penin1214}
\end{figure}

According to the halo models (see Fig.~\ref{fig:cib_source_dist}), the CIB monopole has a broad redshift distribution, peaks at $z=2$ and extends to $z \sim 4$.
However, as we explain above, lensing is reconstructed from the CIB fluctuations and not the CIB monopole. 
The relevant redshift distribution is therefore that of the CIB power spectrum. It is much more skewed towards low redshift, especially for $\ell \lesssim 1000$.
Finally, we compare the measured Planck CIB power spectrum at 545GHz to the halo model prediction in Fig.~\ref{fig:cl_cib_545_planck_penin1214}.

\section{Effect of lensing on the CIB power spectrum}
\label{sec:ps}

The halo model we implement in this paper \cite{2014MNRAS.439..143P} already includes the effect of strong lensing on the CIB. 
Indeed, this halo model relies on the galaxy flux distribution function from \cite{2012ApJ...757L..23B}
, which includes the effect of highly magnified dusty star forming galaxies on the observed flux distribution (see Fig.~3 in \cite{2012ApJ...757L..23B}).
In this section, we focus instead on the effect of weak lensing on the CIB power spectrum.

For a source image at a given redshift, gravitational lensing deflects light  by an angle\footnote{Again, we follow the optical lensing sign convention $T(\n) = T^0(\n - \mbf{d})$, $d=\mbf{\nabla}\phi$ and $\kappa=\frac{1}{2}\nabla^2 \phi$ from \cite{2001PhR...340..291B}, as opposed to the CMB lensing sign convention $T(\n) = T^0(\n + \mbf{d})$, $d=\mbf{\nabla}\phi$ and $\kappa=-\frac{1}{2}\nabla^2 \phi$ from \cite{2006PhR...429....1L}.}
$\mbf{d} = \mbf{\nabla} \phi$.
As described in \cite{2001PhR...340..291B}, the observed map $T$ (in units of temperature, surface brightness, number counts or any other quantity that gets lensed) at final position $\n$ is equal to the unlensed source map $T^0$ at position $\n - \mbf{d}$ :
$T(\n) = T^0(\n - \mbf{d})$.
Taylor expanding this equation in $\mbf{d} = \mbf{\nabla}\phi$ thus allows one to compute the lensed power spectrum \cite{2006PhR...429....1L}.

We approximate the CIB as a single source at redshift $z=2$. In this single-source approximation, the lensed power spectrum can be computed to `2-loop' order (i.e. to second order in $C^\phi$) as \cite{2004NewA....9..173C}:
\beq
\bal
C_\ell &= 
C^0_\ell 
\left[ 1 - \ell^2 D^2 + \frac{1}{2} \ell^4 D^4 \right] \\
&+ \int \frac{d^2L}{(2\pi^2)} 
C^\phi_L C^0_{\vl-\vL} 
\left[ \vL\cdot\left( \vl-\vL \right)  \right]^2
\left[ 1 - |\vl-\vL|^2 D^2 \right]\\
&+\frac{1}{2} \int \frac{d^2L_1d^2L_2}{(2\pi)^4}
C^\phi_{L_1} C^\phi_{L_2} C^0_{\vl - \vL_1 - \vL_2}
\left[ \vL_1\cdot\left( \vl-\vL_1-\vL_2 \right) \;
\vL_2\cdot\left( \vl-\vL_1-\vL_2 \right) \right]^2 \\
&+\mathcal{O}\left( (C^\phi)^3 \right),\\
\eal
\label{eq:lensed_power}
\eeq
where 
$D^2 \equiv \int \frac{dL}{4\pi} L^3 C^\phi_L$
is half of the mean squared deflection.
The result is shown in Fig.~\ref{fig:cl_cib_545_planck_penin1214}: weak lensing  changes the CIB power spectrum by less than a percent for $\ell < 10^4$.
As pointed out in \cite{2006PhR...429....1L}, 
this is expected when the unlensed power spectrum is close to featureless, resembling a power law.
A more precise calculation of the lensed CIB power spectrum is possible, taking account of the extended redshift distribution of the CIB emission \cite{2004NewA....9..173C}. However, given the small size of the effect, such a calculation appears unnecessary.
Such a small change in the CIB power spectrum seems extremely difficult to detect, as it would require a prior knowledge of the unlensed CIB to better than one percent.
However, although weak lensing does not change the mean CIB power spectrum in a detectable way, it causes detectable spatial modulations of the power spectrum, as we show below.

\section{Quadratic lensing estimator}
\label{sec:QE}

While lensing has little effect on the power spectrum, it introduces statistical inhomogeneities in the lensed maps, which can be used to reconstruct the lensing potential.
Here we first follow the derivation and notations of the quadratic estimator of \cite{2002ApJ...574..566H}, and then generalize the formalism to include non-Gaussian source distributions.
Expanding $T^0$ at position $\n - \mbf{\nabla}\phi$ to linear order in $\phi$ yields
\beq
T(\n) = T^0(\n) - \mbf{\nabla}\phi \cdot \mbf{\nabla} T^0 + \mathcal{O}(\phi^2) ,
\eeq
or in Fourier space:
\beq
T_\vl = T^0_\vl + \int \frac{d^2\vL}{(2\pi)^2} \; \vL \cdot (\vl - \vL) \; \phi_\vL T^0_{\vl - \vL} + \mathcal{O}(\phi^2).
\eeq
The unlensed map is assumed to be statistically homogenous (or isotropic for the curved sky), which translates into uncorrelated Fourier modes:
\beq
\langle T^0_{\vl} T^0_{\vl'} \rangle = (2\pi)^2 \delta^D_{\vl+\vl'} C^0_l ,
\eeq
where $\delta^D$ is the Dirac delta distribution.
In contrast, lensing couples Fourier modes of the unlensed map, thus breaking this statistical homogeneity:
\beq
\langle T_{\vl} T_{\vL-\vl} \rangle
_\text{fixed $\phi$}
=
\phi_\vL
 \;
\underbrace{
\left(-\vL\right) \cdot
\left[ 
\vl C^0_{\vl}
+
 (\vL - \vl) C^0_{\vL - \vl}
 \right]}_
{f_{\vl, \vL-\vl}} \;
+ \;
\mathcal{O}(\phi^2 C^0) .
\label{eq:break_stat_isotropy}
\eeq
From the last equation, we see that for a fixed realization of the lensing potential, the quadratic quantity 
\beq
\hat{\phi}_{\vL, \vl} \equiv \frac{T_{\vl} T_{\vL-\vl}}{ f_{\vl, \vL-\vl}}
\eeq
 is an estimator for $\phi_\vL$, unbiased to linear order in $\phi$:
\beq
\left \langle 
\frac{T_{\vl} T_{\vL-\vl}}
{f_{\vl, \vL-\vl}}
 \right \rangle_\text{fixed $\phi$}
 =
 \phi_\vL
 \;+\; \mathcal{O}(\phi^2).
\eeq
Note that $f_{\vl, \vL-\vl} = f_{\vL-\vl, \vl}$ and $\hat{\phi}_{\vL, \vl} = \hat{\phi}_{\vL, \vL-\vl}$ by symmetry, and $\hat{\phi}_{\vL, \vl}^\star = \hat{\phi}_{-\vL, -\vl}$.
We thus obtain many estimators of $\phi_\vL$, by fixing $\vL$ and varying $\vl$ in  $\hat{\phi}_{\vL, \vl}$.
These may be combined into the unbiased (to first order in $\phi$) minimum-variance quadratic estimator for $\phi_\vL$, with inverse-variance weighting.
This minimum-variance weighting is different for a Gaussian and non-Gaussian unlensed map, as we now present.

\subsection{Gaussian unlensed map}

First, consider the case of a Gaussian unlensed map $T^0$, as for the CMB. For a fixed realization of the lensing field, the lensed map is then Gaussian too, and we get:
\beq
\langle \hat{\phi}_{\vL, \vl} \hat{\phi}_{\vL^\prime, \vl^\prime}^\star \rangle
_\text{fixed $\phi$}
=
(2\pi)^2 \delta^D_{\vL-\vL^\prime}
\frac{(2\pi)^2 \left( \delta^D_{\vl^\prime-\vl} + \delta^D_{\vl^\prime- (\vL-\vl)} \right)}{2}
\;
\underbrace{
\frac{
2 C^\text{total}_\vl C^\text{total}_{\vL-\vl} 
}
{f_{\vl, \vL-\vl}^2}
.}
_{\equiv \sigma^2_{\vL, \vl}}
\eeq
Note that we have fixed the lensing realization and marginalized over the unlensed map. 
This equation implies that the various estimators $\hat{\phi}_{\vL, \vl}$ are uncorrelated, and the minimum-variance unbiased (to linear order in $\phi$) quadratic estimator takes the simple form:
\beq
\hat{\phi}_\vL
=
\frac{
\int \frac{d^2\vl}{(2\pi)^2} \; \hat{\phi}_{\vL, \vl} / \sigma^2_{\vL, \vl}
}
{
\int \frac{d^2\vl}{(2\pi)^2} \; 1 / \sigma^2_{\vL, \vl}
}.
\label{eq:gaussian_weights}
\eeq
As usual, inverse-variance weighting optimally weights modes according to whether they are signal or noise-dominated.
It further upweights temperature modes $\mbf{\ell}$ that are more aligned with the lensing mode $\mbf{L}$, since these are most affected by lensing.
The optimal quadratic estimator can be recast into the familiar form \cite{2010GReGr..42.2197H}:
\beq
\hat{\phi}_\vL
=
N_\vL \; 
\int \frac{d^2\vl}{(2\pi)^2} \;
\underbrace{i\vL \cdot}_\text{divergence}
\left[
\underbrace{\frac{T_{\vL-\vl}}{C_{\vL-\vl}^\text{total}}}_{\substack{\text{inverse-var.} \\ \text{weighted map}}}
\times
\underbrace{\frac{C^0_\vl}{C^\text{total}_\vl} \ i\vl \ T_\vl }_{\substack{\text{Wiener-filtered} \\ \text{gradient map}}}
\right]
.
\eeq
We recover the fact that the quadratic estimator for a Gaussian unlensed map is the divergence of the real-space product of the inverse-variance weighted unlensed map by its Wiener-filtered gradient.
The reconstruction noise is then:
\beq
N_\vL = \left[ \int \frac{d^2\vl}{(2\pi)^2} \; \frac{1}{\sigma^2_{\vL, \vl}} \right]^{-1}
= 
\left[
\int \frac{d^2\vl}{(2\pi)^2} \;
\frac{f_{\vl, \vL-\vl}^2}
{2C^\text{total}_\vl C^\text{total}_{\vL-\vl}}
\right]^{-1}
.
\label{eq:n0}
\eeq
In this Gaussian case, the quadratic temperature pairs are uncorrelated, so they each add independent information.
This will no longer be the case for a non-Gaussian unlensed map, as we now show.

\subsection{Non-Gaussian unlensed map}
\label{subsec:NG}

If the unlensed map is non-Gaussian, it may have a non-zero trispectrum $\mathcal{T}^0$, defined as the connected component of the four-point function: $\langle T^0_{\vl_1} T^0_{\vl_2} T^0_{\vl_3} T^0_{\vl_4} \rangle_c \equiv (2 \pi)^2 \delta^D_{\vl_1 + \vl_2 + \vl_3 + \vl_4} \mathcal{T}^{0}_{\vl_1, \vl_2, \vl_3, \vl_4} $. 
This both increases the statistical error of the quadratic lensing estimators, and correlates them\footnote{In principle, the right-hand side should also include the self-lensing bispectrum described in Sec.~\ref{subsec:self-lensing}. In practice, neglecting this term only increases the lensing noise, but does not introduce further bias.}
:
\beq
\langle \hat{\phi}_{\vL, \vl} \hat{\phi}_{\vL^\prime, \vl^\prime}^\star \rangle
_\text{fixed $\phi$}
=
(2\pi)^2 \delta^D_{\vL-\vL^\prime}
\;
\underbrace{
\left[
\frac{
(2\pi)^2 \left( \delta^D_{\vl^\prime-\vl} + \delta^D_{\vl^\prime- (\vL-\vl)} \right)
C^\text{total}_\vl C^\text{total}_{\vL-\vl} 
+\mathcal{T}^0_{\vl, -\vl^\prime, \vL-\vl, \vl^\prime-\vL}
}
{f_{\vl, \vL-\vl} \; f_{\vl^\prime, \vL-\vl^\prime}}
\right]}
_{\equiv \left(\Sigma_{\vL}\right)_{\vl, \vl^\prime}}
.
\eeq
Because the estimators $\hat{\phi}_{\vL, \vl}$ are now correlated for different $\vl$, the inverse-variance weighting should  include the non-diagonal covariance matrix $\left(\Sigma_{\vL}\right)_{\vl, \vl^\prime}$: 
\beq
\hat{\phi}_\vL
=
\frac{
\int \frac{d^2\vl}{(2\pi)^2} \; \hat{\phi}_{\vL, \vl}\;  
\left[
\int \frac{d^2\vl^\prime}{(2\pi)^2} \; \left( \Sigma_{\vL} \right)^{-1}_{\vl, \vl^\prime}
\right]
}
{
\int \frac{d^2\vl d^2\vl^\prime}{(2\pi)^4} \; \left( \Sigma_{\vL} \right)^{-1}_{\vl, \vl^\prime}
},
\eeq
where the inverse-covariance matrix in the continuum limit is defined by
\beq
\int \frac{d^2\vl}{(2\pi)^2} \; 
\left( \Sigma_{\vL} \right)^{-1}_{\vl_1, \vl}
\left( \Sigma_{\vL} \right)_{\vl, \vl_2}
=
(2\pi)^2
\delta^D_{\vl_1-\vl_2}.
\eeq
The associated reconstruction noise power spectrum is
\beq
N_\vL
=
\left[
\int \frac{d^2\vl d^2\vl^\prime}{(2\pi)^4} \; \left( \Sigma_{\vL} \right)^{-1}_{\vl, \vl^\prime}
\right]^{-1}
.
\label{eq:N0_nongausssource_nongaussweights}
\eeq
As a result, the noise in the lens reconstruction is enhanced by the presence of the trispectrum.
This lensing noise constitutes a bias in the auto-power spectrum of the reconstructed lens field, as for the Gaussian case.

To add some intuition and aid comparing to the Gaussian case, we evaluate Eq.~\eqref{eq:N0_nongausssource_nongaussweights} in the simple case of a pure Poisson unlensed map.
That is, both the power spectrum $C^0$ and the trispectrum $\mathcal{T}^0$ are white, i.e. independent of multipole.
In this particular case, the lensing noise in units of $\kappa$ becomes (see App.~\ref{app:N0_nongausssource} for a complete derivation):
\beq
\bal
N_\vL^\kappa
&=
\underbrace{\frac{2\pi}{N_\text{modes}}}
_{\text{Gaussian noise}}
+
\underbrace{\frac{\mathcal{T}^0}{4 (C^0)^2}}
_{\text{non-Gaussian correction}},
\eal
\label{eq:lensing_noise_poisson}
\eeq
where $N_\text{modes} \equiv 4\pi \int \frac{d^2\vl}{(2\pi)^2} = \ell_\text{max}^2 - \ell_\text{min}^2$.
This simple formula provides several key insights.
The first term is the Gaussian contribution. 
This term goes to zero as the number of observed multipoles increases: if the modes are Gaussian, each of them adds independent lensing information, and the number of modes is the only limit to the reconstruction precision.
The second term is the additional lensing noise due to the non-Gaussianity of the unlensed map. 
It does not decrease as the number of observed modes increases.
Thus, in the limit of an infinite number of modes, the Gaussian term tends to zero, and the total noise is entirely determined by the non-Gaussian term.
Another important insight is the following.
For the purpose of lens reconstruction, a small ratio between the trispectrum and the squared power spectrum, i.e. $\frac{\mathcal{T}^0}{(C^0)^2} \ll 1$, is not sufficient for the field to be considered Gaussian.
One needs the trispectrum to be small compared to the squared power spectrum divided by the number of modes $N_\text{modes}$:
$\frac{\mathcal{T}^0}{(C^0)^2/N_\text{modes}} \ll 1$.
This is a more stringent condition, and one that depends on the experiment.
As we show below, it is satisfied by Planck but not by CCAT-Prime.

\subsection{Applying Gaussian weights to a non-Gaussian map}

In order to avoid inverting the large covariance matrix, or if the unlensed trispectrum is not well-known,
one may apply the quadratic estimator with Gaussian weights to a non-Gaussian map.
If the non-Gaussianity is large, i.e. $\frac{\mathcal{T}^0}{(C^0)^2/N_\text{modes}} \not\ll 1$, the Gaussian weighting is suboptimal: 
it may increase significantly the reconstruction noise, but does not introduce bias.
When applying the Gaussian weights of Eq.~\eqref{eq:gaussian_weights} to a non-Gaussian map, the lensing noise power spectrum is no longer given by Eq.~\eqref{eq:N0_nongausssource_nongaussweights}, but instead: 
\beq
N_\vL = 
\frac{
\int \frac{d^2\vl d^2\vl^\prime}{(2\pi)^4} \; \left(\Sigma_{\vL}\right)_{\vl, \vl^\prime} / \sigma^2_{\vL, \vl}\sigma^2_{-\vL, \vl^\prime}
}{
\left[ \int \frac{d^2\vl}{(2\pi)^2} \; 1 / \sigma^2_{\vL, \vl} \right]^2
}
= 
N^{0}_L + \mathscr{T}_L^{0} \ , 
\label{eq:N0_nongausssource_gaussweights}
\eeq
where the Gaussian reconstruction noise $N^{0}_L$ is the same as in Eq. \eqref{eq:n0},
\beq
N^{0}_L = \left[ \int \frac{d^2\vl}{(2\pi)^2} \; \frac{1}{\sigma^2_{\vL, \vl}} \right]^{-1}
= 
\left[
\int \frac{d^2\vl}{(2\pi)^2} \;
\frac{f_{\vl, \vL-\vl}^2}
{2C^\text{total}_\vl C^\text{total}_{\vL-\vl}}
\right]^{-1}
\ ,
\eeq
and the additional contribution due to the unlensed source trispectrum is
\beq
\mathscr{T}_L^{0} 
= 
(N^{0}_{L})^2 
\int \frac{d^2\vl d^2\vl^\prime}{(2\pi)^4} \ 
\frac{f_{\vl, \vL - \vl} }{2C^\text{total}_\vl C^\text{total}_{\vL-\vl}} 
\frac{f_{\vl', -\vL - \vl'} }{2C^\text{total}_{\vl'} C^\text{total}_{-\vL-\vl'}}  \ 
\mathcal{T}^0_{\vl, \vL - \vl, \vl', -\vL - \vl'}.
\label{eq:T0}
\eeq
As expected, the reconstruction noise is enhanced by the unlensed trispectrum relative to the Gaussian case.

In particular, for a white trispectrum (but an arbitrary power spectrum), this non-Gaussian noise simplifies to
\beq
\mathscr{T}_L^{0} 
= 
(N^{0}_{L})^2 
\left[
\int \frac{d^2\vl}{(2\pi)^2} \ 
\frac{f_{\vl, \vL - \vl} }{2C^\text{total}_\vl C^\text{total}_{\vL-\vl}}  
\right]^2
\mathcal{T}^0.
\label{eq:T0_whiteT}
\eeq
In Sec.~\ref{sec:forecasts}, we evaluate this expression to quantify the effect of the shot noise trispectrum on the CIB lensing estimator.
If the power spectrum and trispectrum are both white, e.g., in the shot noise regime of the CIB, Eq.~\eqref{eq:T0_whiteT} further simplifies to $\mathscr{T}_L^{0} = \mathcal{T}^0/(L^4C^2)$ (App.~\ref{app:N0_nongausssource}).
In this very special case, despite the potentially large non-Gaussianity, we recover the corresponding result for the non-Gaussian weights. 
Indeed, if both the power spectrum and the trispectrum are white, then all temperature modes are equivalent, and will be weighted equally by both the Gaussian and non-Gaussian weights.

\subsection{CIB trispectrum: halo model prediction and map-based measurement}

As we have shown, the unlensed CIB trispectrum contributes to the lensing noise.
Furthermore, as we shall demonstrate shortly, 
the quadratic CIB lensing estimator best reconstructs the low lensing multipoles $L$, and relies mostly on the high temperature multipoles $\ell$.
In other words, 
the relevant trispectrum configuration is the collapsed one: $\mathcal{T}^\text{CIB}_{\vl, \vL-\vl, \vl ', -\vL - \vl '}$, with $L \ll \ell$.
In what follows, we therefore focus exclusively on this configuration.
This configuration can be measured from the data without implementing a complete trispectrum estimator (see App.~\ref{app:trispectrum_measurement}),
and it also simplifies the halo model calculations (see App.~\ref{app:cib_halo_model}).

\begin{figure}[h]
\centering
\includegraphics[width=0.49\columnwidth]{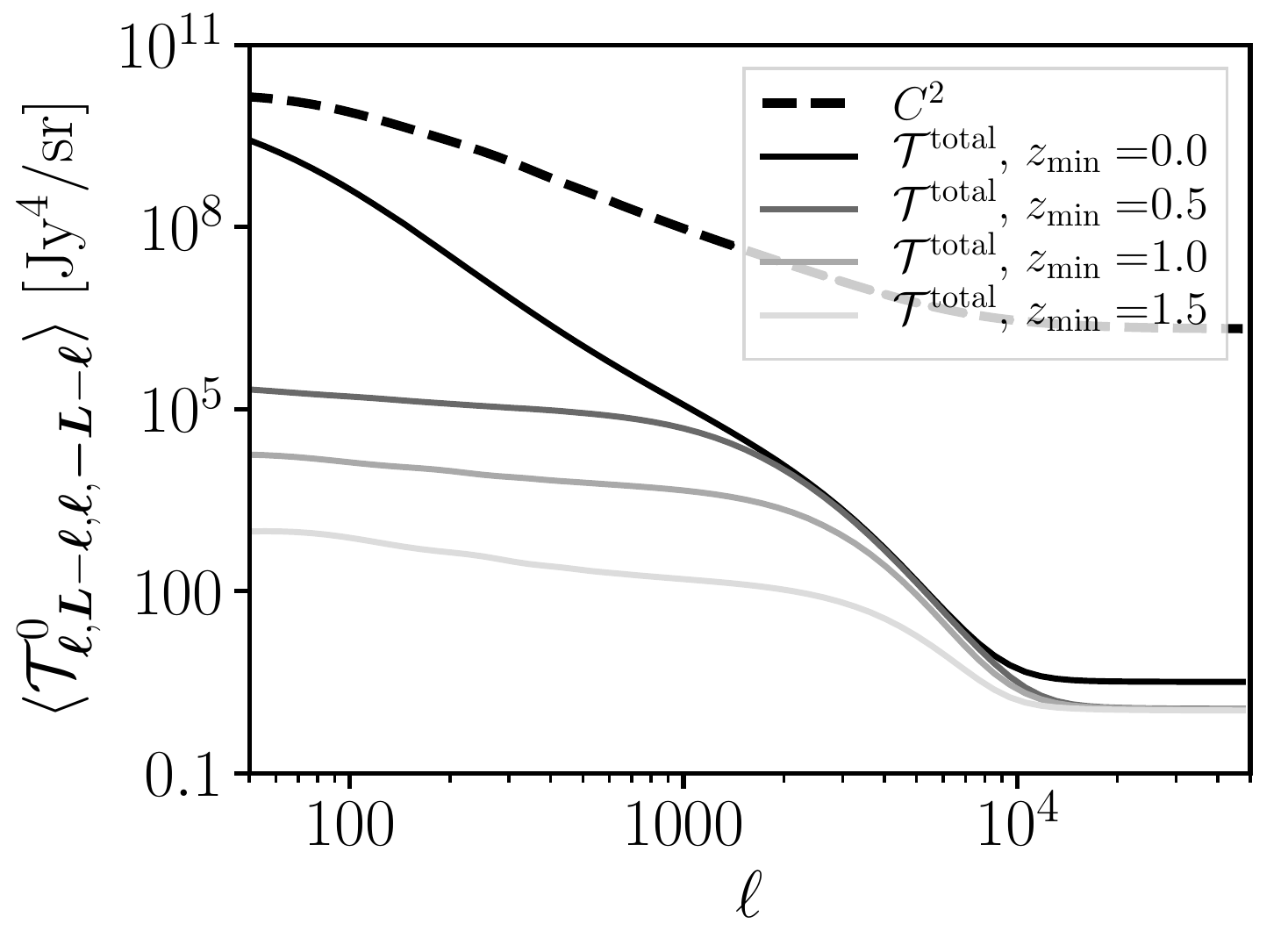}
\includegraphics[width=0.49\columnwidth]{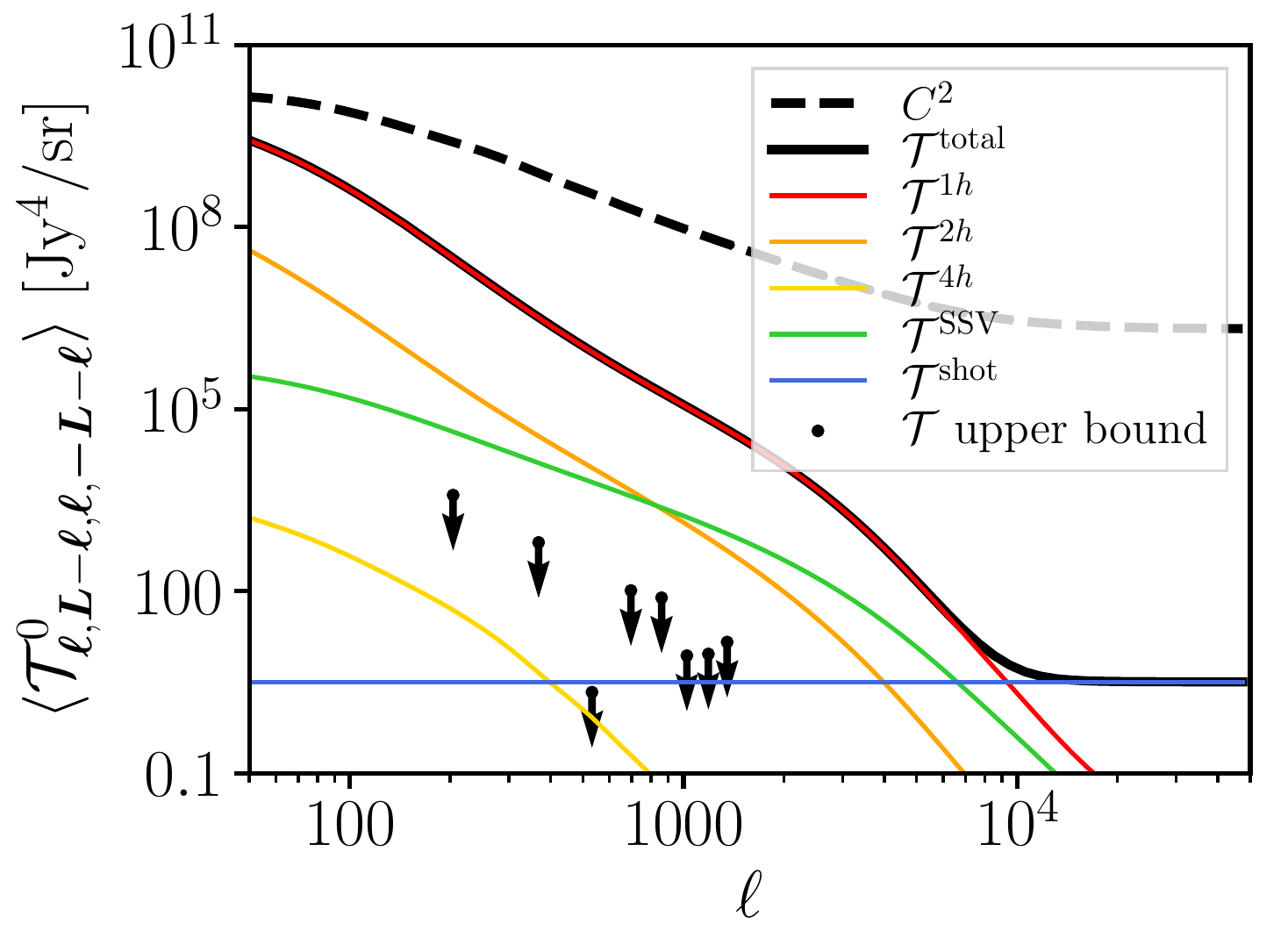}
\caption{
Measure of the CIB non-Gaussianity relevant for the lensing noise power spectrum:
comparison between the angle-averaged collapsed CIB trispectrum (solid black line) and the squared CIB power spectrum (dashed black line)\\
\textbf{Left panel:} We show the large dependence of the halo model trispectrum on the minimum redshift cut $z_\text{min}$
in the halo model trispectrum Eq.~\eqref{eq:trispectrum_halo_model}.
We vary this value from $0$ to $1.5$ and show the corresponding CIB trispectra in shades of grey.\\
\textbf{Right panel:} Keeping $z=0$ as the minimum redshift cut, we compare the various halo model terms (colored lines) to our trispectrum measurement from the Planck GNILC map at 545GHz.
The measurement procedure is presented in App.~\ref{app:trispectrum_measurement} 
We find no significant evidence for a non-zero collapsed trispectrum
and show one-sided $2\sigma$ upper bounds
as downward triangles.
The non-detection of a collapsed trispectrum with the optimal estimator for Planck implies that CIB non-Gaussianity can be safely ignored for the CIB lens reconstruction with Planck.\\
The measurement only extends to $\ell=1500$, and therefore does not directly constrain the higher multipoles. However, the fact that most halo model terms are excluded at low multipoles suggests that only the shot noise trispectrum will be significant at higher multipoles. 
We include it in the Planck and CCAT-Prime forecasts, which rely primarily on these high multipoles.
}
\label{fig:cib_trispectrum}
\end{figure}

In Fig.~\ref{fig:cib_trispectrum}, we show estimates of the CIB trispectrum from the halo model, together with a map-based measurement.
As shown in the left panel, the halo model prediction for the CIB trispectrum appears to be extremely sensitive to the 
low-redshift 
CIB sources, as defined in App.~\ref{app:trispectrum_hm}.
This makes it difficult to reliably estimate the CIB trispectrum from the halo model, given current observables. 
For this reason, we also measured the collapsed trispectrum from the Planck GNILC CIB map at 545GHz. 
The details of the method are presented in App.~\ref{app:trispectrum_measurement}. 
Our measurement from the Planck maps extends to $\ell=1500$.
Over this range of multipoles, we do not detect a collapsed trispectrum but place an upper limit.
This upper limit excludes most halo model components below $\ell=1500$, except for the 4-halo and shot noise terms,
This implies that the `excluded' terms are highly overestimated in the halo model we implemented.
Although our measurement does not constrain the higher multipoles directly, it suggests that the shot noise trispectrum is the only dominant term for $\ell\gtrsim1500$, where most of the lensing information originates.
Indeed, all other halo model terms decrease rapidly with $\ell$, so if their amplitudes were consistent with the upper limit at $\ell=1500$, they would be small compared to the shot noise at $\ell\gtrsim1500$.
As shown in Fig.~\ref{fig:ciblensrecnoise}, the non-Gaussian lensing noise due to the shot noise trispectrum is negligible compared to the Gaussian contribution for Planck, and dominant for CCAT-Prime.
Thus, the Gaussian lensing weights are close to optimal for Planck and likely suboptimal for CCAT Prime.

The halo model term which is least uncertain is probably the shot noise trispectrum, since it comes from measured luminosity functions. As we show below, CIB lensing from CCAT-Prime relies mostly on temperature multipoles above $\ell=10,000$, where the shot noise is the only dominant term. As a result, our forecast for the non-Gaussian lensing noise for CCAT-Prime should be robust.

Since the CIB non-Gaussianity increases the lensing noise for CCAT-Prime,
one natural question is whether the CIB can be `Gaussianized'.
An efficient way of achieving this is to mask rare, bright objects, which contribute little to the power spectrum but a lot to the trispectrum.
For instance, the deeper CCAT-Prime flux cut reduces the shot noise power spectrum by less than $10\%$ (compared to the 350mJy flux cut of Planck),
but lowers the shot noise trispectrum by an order of magnitude.
Masking known infrared sources from galaxy surveys should further reduce the CIB trispectrum, while marginally affecting the CIB power spectrum.

On the other hand, the apparent discrepancy between the halo model and the data also suggests that the higher point functions of the CIB contain useful information to distinguish between different halo models of the CIB, and thus constrain the history of star formation in the universe.

\section{Forecasts}
\label{sec:forecasts}

\begin{figure}[h]
\centering
\includegraphics[width=0.49\columnwidth]{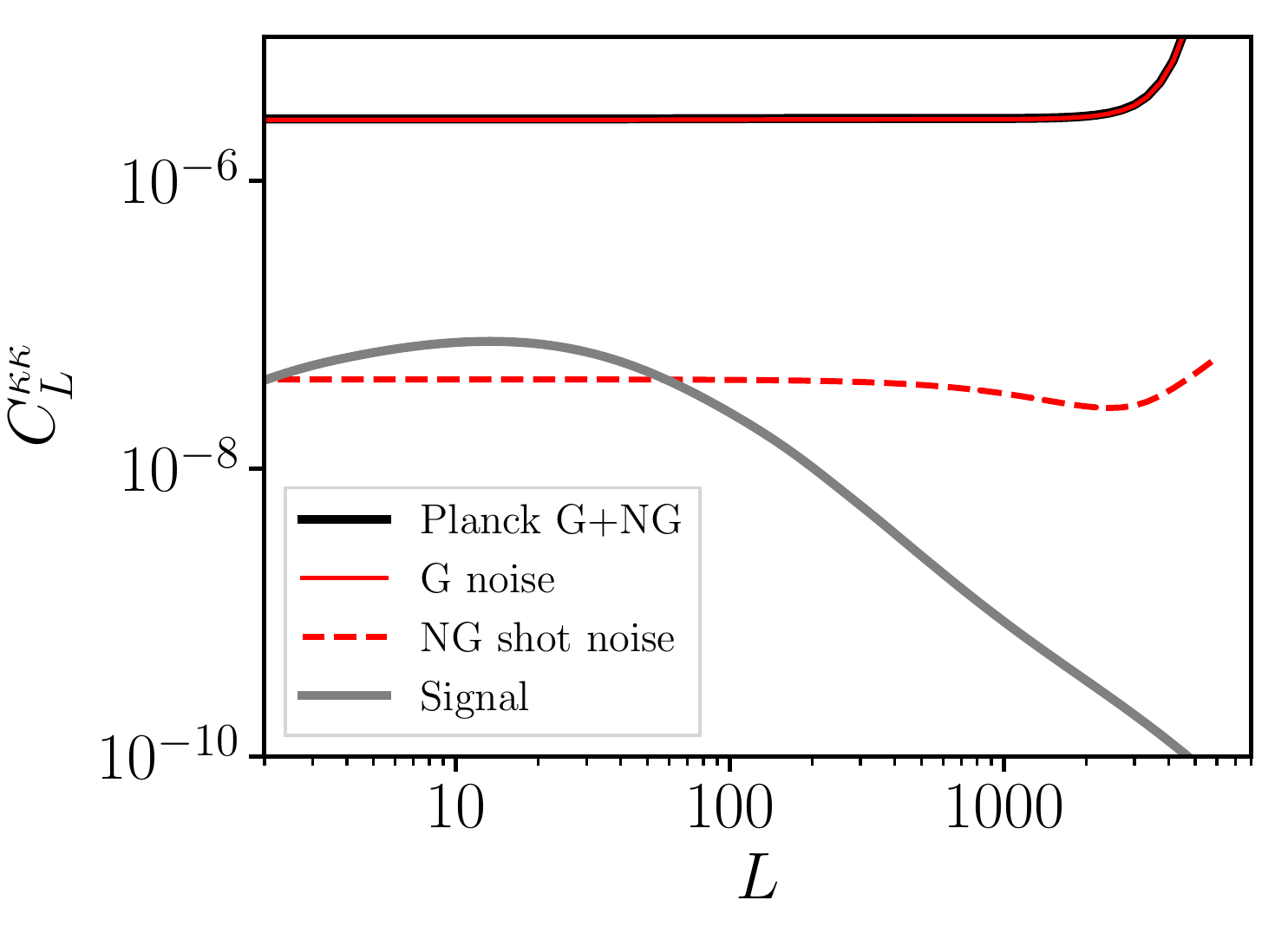}
\includegraphics[width=0.49\columnwidth]{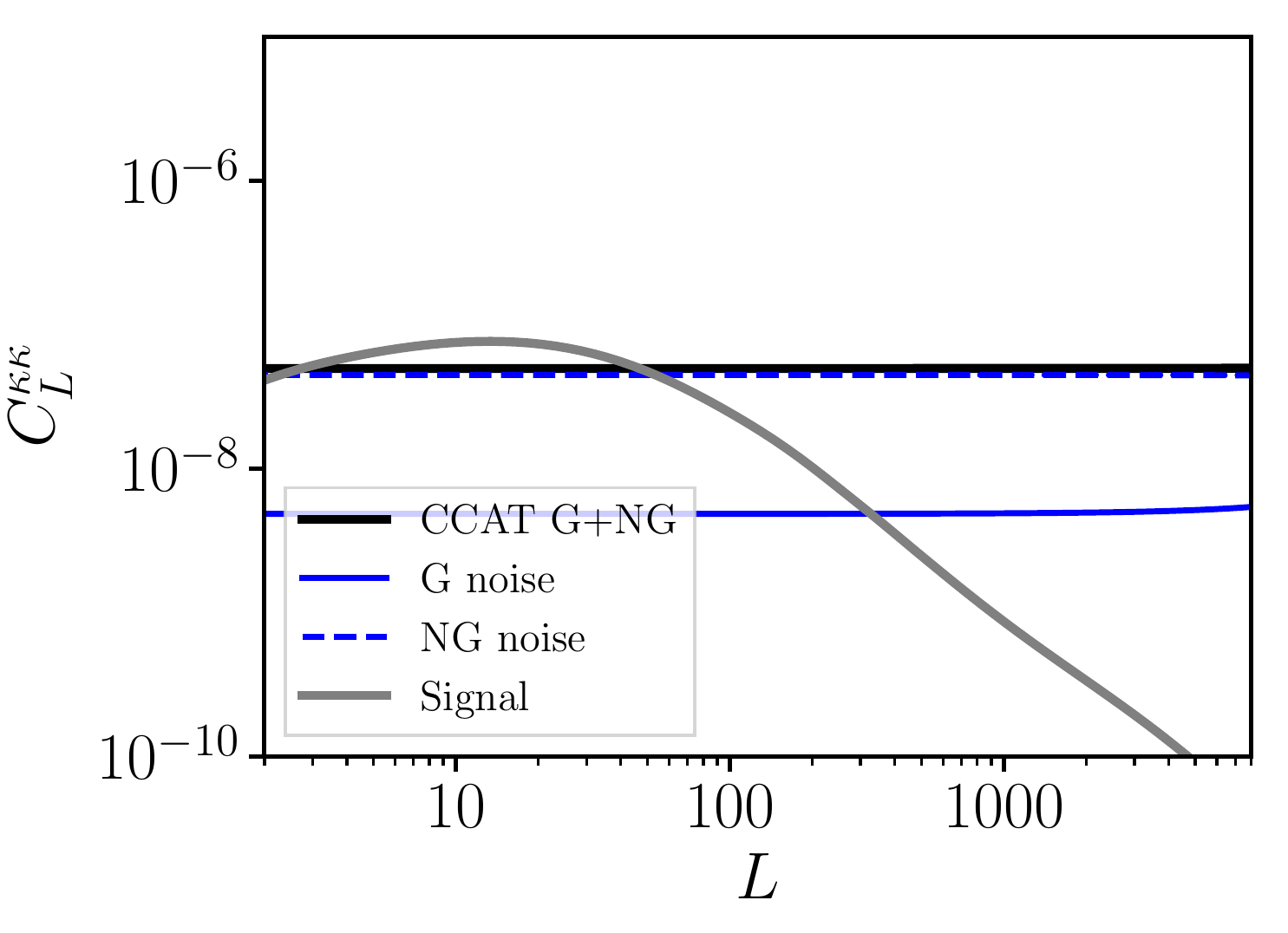}
\caption{
Noise per multipole on the CIB lens reconstruction, using Gaussian weights for the quadratic estimator.
Solid black lines show the lensing signal and the lensing noise. 
The solid red curve shows the lensing noise expected if the CIB were perfectly Gaussian.
The dashed red curve shows the additional noise contribution from the shot noise trispectrum.
\textbf{Left panel:} Planck CIB lensing. 
As explained in the main text, the CIB trispectrum does not alter the lensing noise in a detectable manner, since it is not detected by our collapsed trispectrum estimator.
In particular, the lensing noise due to the CIB shot noise trispectrum is negligible compared to the Gaussian noise.\\
\textbf{Right panel:} CCAT-Prime CIB lensing. 
The lensing noise is dominated by the non-Gaussian contribution, i.e. the CIB shot noise trispectrum.  
The lensing modes between a few and 50 are signal dominated.\\
CIB lensing from CCAT-Prime relies on higher multipoles than for Planck (see Fig.~\ref{fig:lensing_contribution}).
While the shot noise trispectrum is smaller for CCAT-Prime than for Planck, due to the lower flux cut, the CIB power spectrum is also smaller at these higher multipoles. As a result, the non-Gaussian lensing noise is similar for CCAT-Prime and Planck, scaling roughly as $\frac{\mathcal{T}^0}{4 (C^0)^2}$.
}
\label{fig:ciblensrecnoise}
\end{figure}

\begin{figure}[h]
\centering
\includegraphics[width=0.4\columnwidth]{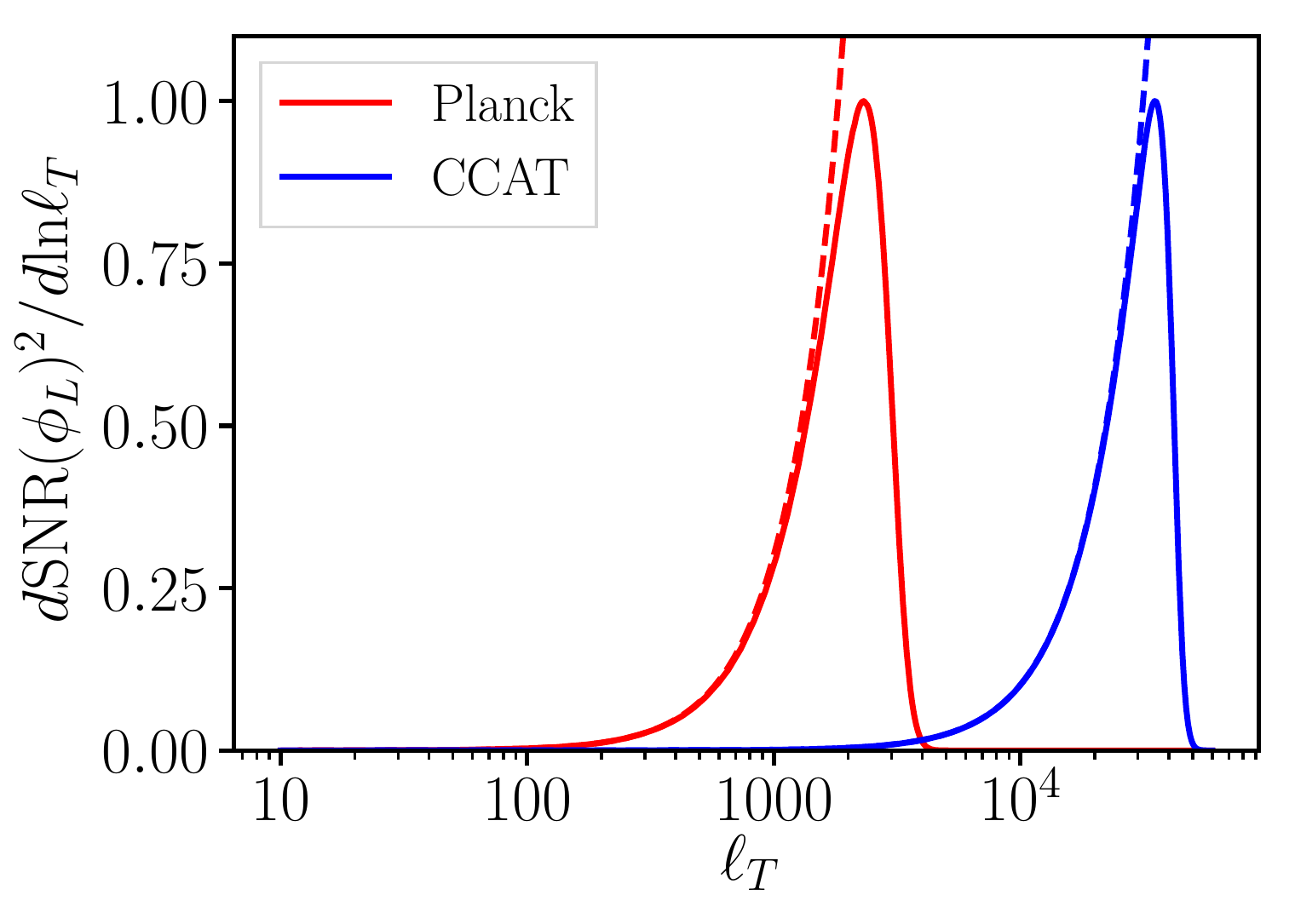}
\caption{
Contribution from the various $\ell$-modes of the CIB temperature map to the lensing potential at $L=100$, for Planck (solid red curve) and CCAT-Prime (solid blue curve).
To show both curves on the same axis, they have been normalized to a maximum of unity.
The dashed lines show the expectation for Planck and CCAT-Prime when including only sample variance mode counting ($N_L^\kappa = 2\pi/N_\text{modes}$ from Eq.~\eqref{eq:lensing_noise_poisson}).
They show that most of the lensing information comes from the highest multipoles that are signal dominated in the CIB map, since they contribute the largest number of modes.
For these high multipoles, the trispectrum can be approximated by the shot noise term.
}
\label{fig:lensing_contribution}
\end{figure}

We forecast the signal-to-noise ratio for the CIB lensing auto and cross-correlations, for Planck and CCAT-Prime.
Our specifications are presented in Tab.~\ref{tab:specs_planck_ccat}.
Because the CIB trispectrum is very uncertain, we assume Gaussian weights for the quadratic estimator for both Planck and CCAT-Prime.
As discussed above, this should be optimal for Planck, but not for CCAT-Prime. 
In both cases, we compute the noise contribution due to the CIB shot noise trispectrum.
The resulting lensing noises for Planck and CCAT-Prime are shown in Fig.~\ref{fig:ciblensrecnoise}, on the left and right panel, respectively.
In Fig.~\ref{fig:lensing_contribution}, we show that the lens reconstruction is dominated by the highest signal-dominated multipoles available in the experiment:
$\ell\sim1000$ for Planck and $\ell\sim10,000$ for CCAT-Prime.
This justifies retaining only the shot noise term in the CIB trispectrum.

\begin{figure}[h]
\centering
\includegraphics[width=0.4\columnwidth]{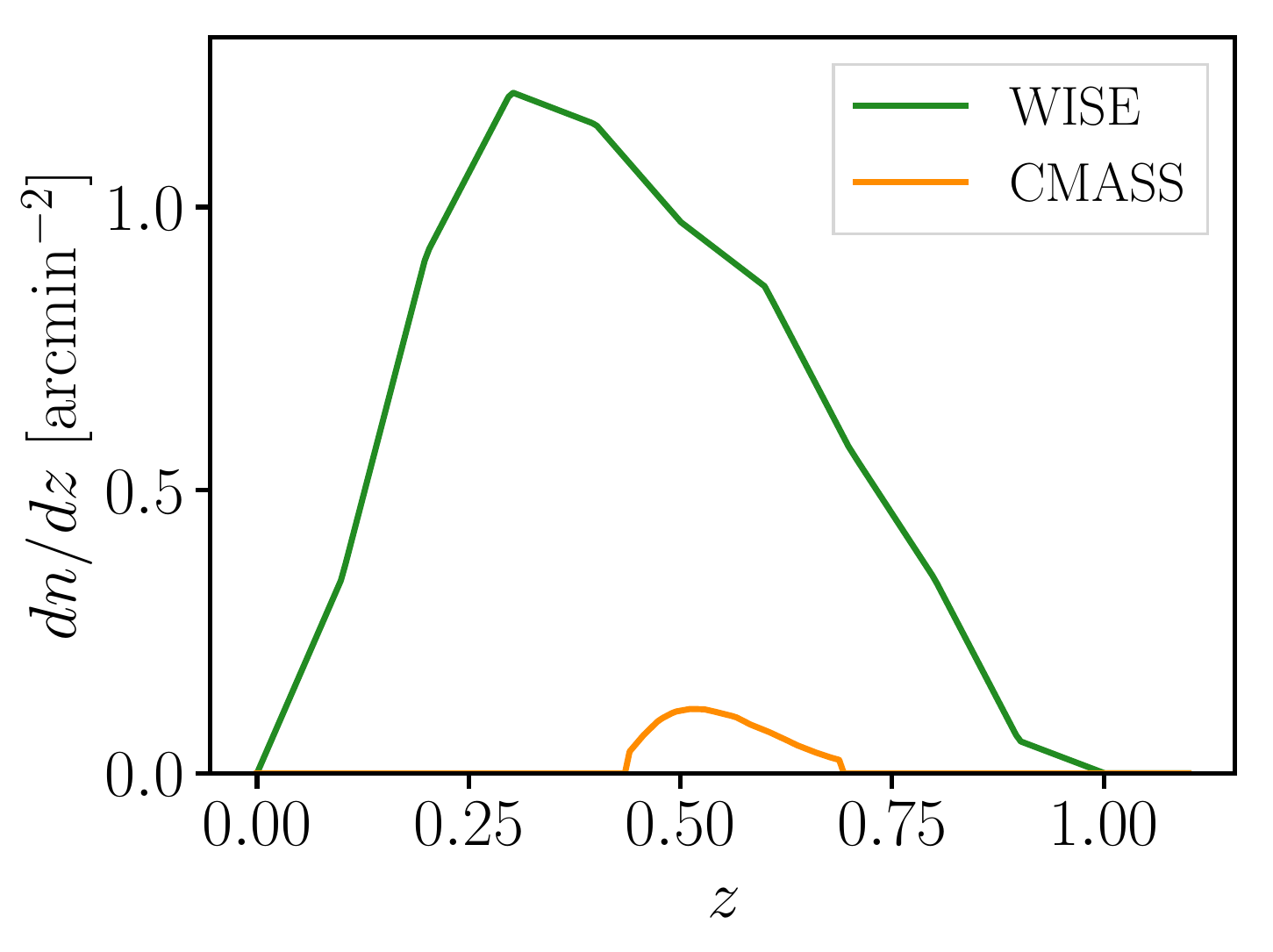}
\caption{
Redshift distribution for WISE (green curve) and CMASS (orange curve). As lens galaxies, these may be correlated with the reconstructed CIB lensing potential.
}
\label{fig:dndz_cmass_wise}
\end{figure}

We then forecast the signal-to-noise ratio on the auto and cross-correlations between CIB lensing and various tracers of the matter density. The results are summarized in Tab.~\ref{tab:forecast_snr}.
For Planck CMB lensing, we consider CMB maps with $5^\prime$ beam FWHM, $60 \ \mu$K$^\prime$ noise in temperature and $60 \sqrt{2}\ \mu$K$^\prime$ noise in $E$ and $B$-mode polarizations at $143$ GHz \cite{2016A&A...594A...1P}. 
We use $\ell_\text{max T} = \ell_\text{max P} = 2000$.
This reproduces the lensing noise measured by the Planck collaboration \cite{2016A&A...594A..15P}.
For CMB lensing from CMB S4, we assume a $1^\prime$ FWHM beam, $1\mu$K$^\prime$ noise in temperature and $\sqrt{2}\mu$K$^\prime$ noise in $E$ and $B$-mode polarization at $143$ GHz \cite{2016arXiv161002743A, 2017PhRvD..95l3512S}, and assume $\ell_\text{max T} = 3000$ and $\ell_\text{max P} = 5000$.
For the CMASS galaxy sample, we assume a galaxy number density of $0.02$ arcmin$^{-2}$ and bias of $2$ over $24\%$ of the sky.
For the WISE galaxy catalog, we assume $0.6$ galaxies per squared arcmin, with a bias of $1.13$, over $40\%$ of the sky.
The redshift distributions of CMASS and WISE are shown in Fig.~\ref{fig:dndz_cmass_wise}.

As shown in Tab.~\ref{tab:forecast_snr}, the auto-power spectrum of CIB lensing is not detectable with existing Planck CIB data. 
However, the cross-correlations of Planck CIB lensing with tracers should be significant, including the cross-correlations with CMB lensing, and galaxy catalogs like WISE or CMASS.
With CCAT-Prime data, CIB lensing will be detectable both in auto and cross-correlation at high significance.
CIB lensing will therefore provide an independent measurement of the amplitude of density fluctuations.
However, the cosmological interpretation of this measurement will be affected by the uncertainty on the redshift distribution of the CIB.
On the other hand, one can use CIB lensing measurements and the current best estimates of cosmological parameters to constrain the CIB redshift distribution.
Moreover, if CIB lensing measurements are combined with galaxy or CMB lensing, one can effectively use distance ratios (also referred to as `cosmography') to determine the CIB redshift distribution, without requiring a cosmological model.
\begin{table}[h!]
\begin{center}
\begin{tabular}{ |l|c|c| } 
\hline
 & Planck $\kappa_\text{CIB}$ & CCAT-Prime-like $\kappa_\text{CIB}$ \\
\hline
Auto-power, $f_\text{sky}=0.4$ & 1 & 37 \\  
Planck $\kappa_\text{CMB}$, $f_\text{sky}=0.4$ & 5 & 28 \\
CMB S4 $\kappa_\text{CMB}$, $f_\text{sky}=0.4$ & 12 & 80 \\
WISE, $f_\text{sky}=0.4$ & 18 & 129 \\  
CMASS, $f_\text{sky}=0.24$ & 6 & 40 \\
\hline
\end{tabular}
\end{center}
\caption{Forecasts for the signal-to-noise ratio of each auto/cross-correlation. 
These values of signal-to-noise include the Gaussian and non-Gaussian lensing noise, as well as the cosmic variance.
They thus correspond to the final error on the measured amplitude of the signal. 
The signal-to-noise values for detection alone, i.e. without the cosmic variance, are therefore slightly larger than those quoted in this table.
These signal-to-noise are statistical only, and therefore do not include the biases discussed in Sec.~\ref{sec:biases}.
}
\label{tab:forecast_snr}
\end{table}

\section{Biases to the auto and cross correlation}
\label{sec:biases}

\subsection{Lens reconstruction biases: overview}

In addition to the usual noise biases present in CMB lensing reconstruction \cite{2003PhRvD..67l3507K, 2011PhRvD..83d3005H}, 
the non-Gaussian nature of CIB emission over a broad redshift range introduces new forms of bias. 
These affects both the $\hat{\phi}$ power spectrum and its cross-correlation with tracers in slightly distinct ways.

\begin{itemize}

\item \textbf{Standard noise biases}

It is well-known that the power spectrum of the standard quadratic estimator $\hat{\phi}$ is subject to noise biases, even in the case of Gaussian source and lensing potential.  Schematically we can write 
\beq
\langle C_L^{\hat{\phi} \hat{\phi}} \rangle =  N_L^{(0)} + C_L^{\phi \phi} + N_L^{(1)} + \mathcal{O}([C_L^{\phi \phi}]^2)
\eeq
The $N_L^{(0)}$ term is present even when applying the quadratic estimator to a Gaussian unlensed source field, and represents the disconnected part of the temperature four-point function.  
Following the convention in the literature, we denote the noise biases by $N^{(n)}$, where $n$ is the power of $C_L^{\phi \phi}$ appearing explicitly in the evaluation of the bias term.

At lowest order, the \textit{connected} part of the $\hat{\phi}$ power spectrum expansion contains the true signal $C_L^{\phi \phi}$, and another contraction $N_L^{(1)}$ at the same order (but typically smaller than the signal, at least for CMB lensing) \cite{2003PhRvD..67l3507K, 2011PhRvD..83d3005H}. 
At higher order, additional biases $N_L^{(2)}$, $N_L^{(3)}$, etc. may become important for small-scale or low-noise experiments.
In cross correlation with tracers, 
$N_L^{(2)}$ is the lowest order bias present \cite{2011PhRvD..83d3005H}.

The higher order biases are partially due to the first order Taylor expansion used in deriving the quadratic estimator:
\beq
T(\hat{\mbf{n}}) = T^0(\hat{\mbf{n}} - \mbf{\nabla}\phi) \approx T^0(\hat{\mbf{n}}) - \mbf{\nabla}\phi \cdot \mbf{\nabla} T^0.
\eeq
This linear truncation of the expansion is inexact, and misses all the terms $\mathcal{O}(\phi^n)$ for $n\geq2$. As a result, the quadratic estimator $\hat{\phi}$ is biased by terms $\mathcal{O}(\phi^2)$. 
An estimator including higher order Taylor expansion terms, or a maximum likelihood estimator where the lensed field contains the correct pixel to pixel remapping by lensing, would avoid some part (but not all) of the biases $N_L^{(1)}$, $N_L^{(2)}$, etc.

\item \textbf{Non-Gaussian unlensed map $T^0$}

Each noise term $N_L^{(n)}$ involves a 4-point function of the unlensed temperature map,  and is therefore enhanced by the unlensed trispectrum. Schematically, when using Gaussian weights,
$
N_L^{(n)} \longrightarrow N_L^{(n)} + \mathscr{T}_L^{(n)},
$
where $\mathscr{T}_L^{(n)}$ is an integral over the unlensed trispectrum $\mathcal{T}^{0}$ weighted by $2n$ powers of $\phi$. 
The $n=0$ case is given by Eq.~\eqref{eq:T0}.

As discussed previously, the lowest order bias in cross correlation is $N^{(2)}$. This makes the cross-correlation more robust to uncertainties in the modeling of the trispectrum, which can be hard to disentangle from the lensing contribution.

\item \textbf{Non-Gaussian lensing potential $\phi$}

As the lensing potential is sourced by  large-scale structure, it is affected by the non-linear evolution under gravity. This makes the lensing potential non-Gaussian on small scales. 
A consequence is the appearance of new bias terms $N^{(j)}$ that involve both odd and even powers of $\phi$ (as opposed to even powers only if $\phi$ is Gaussian). 
For example, terms like $N^{(3/2)}$ or $N^{(5/2)}$ appear, and involve the lensing potential bispectrum and 5-point function respectively \cite{2016PhRvD..94d3519B}. The other biases $N^{(n)}$ for integer values of $n$ are also enhanced by the connected $(2n)$-point function of $\phi$. For example, the $N^{(2)}$ bias is enhanced by the 4-point function of $\phi$. These biases also appear in CMB lensing and are subject to active research.
\item \textbf{Self-lensing}

When the source distribution is broad in redshift, as for the CIB, the lower redshift sources can act as lenses for the higher redshift sources, and therefore introduce a bias in both the auto and cross correlation. 
We call this effect self-lensing.
This is analogous to the dominant contribution of some of the foreground-induced biases to CMB lensing \cite{2014ApJ...786...13V, 2014JCAP...03..024O, 2017arXiv170506751F}, in which part of the foregrounds leaks into the temperature map.  
Given the importance of this term, which affects both the auto and cross correlation, we explore it in details in the next section.

\item \textbf{Foreground contamination}

Analogously to CMB lensing, the imperfect removal of foregrounds such as Galactic dust or extragalactic tSZ can lead to biases in the lens reconstruction \cite{2014ApJ...786...13V, 2014JCAP...03..024O, 2017arXiv170506751F}.  
While the tSZ signal can be accurately cleaned due to its unique frequency dependence, Galactic dust is highly degenerate with the CIB, especially at lower frequencies, and residual amounts can lead to a bias for the lensing reconstruction. 

\item \textbf{Magnification bias} 

Magnification bias occurs in galaxy catalogs with a fixed detection threshold.
In magnified regions, galaxies appear brighter, and therefore more galaxies will be detected. 
This effect partially compensates the expected dilution of the galaxy number density due to magnification. 

However, as surface brightness is conserved in lensing, intensity maps with no detection threshold, e.g., in CIB or 21cm intensity mapping \cite{2013PhRvD..87f4026H}, do not suffer from magnification bias.
In practice, when masking bright point source in a realistic CIB analysis, some magnification bias is introduced. Because point sources are relatively rare and have a small effect on the power spectrum (compared to the trispectrum), we expect this bias to be negligible.

\end{itemize}

\subsection{Self-lensing contribution} 
\label{subsec:self-lensing}

The CIB self-lensing bias
 is simpler to evaluate in cross-correlation with tracers than in auto-correlation. 
 We discuss them separately in this section.\\

{\bf Cross-correlation} 

Suppose that we cross-correlate the reconstructed $\hat{\phi}$ with a low-redshift tracer $\delta_g$. 
For concreteness, we consider the cross-correlation with galaxies, but the same formalism applies to any tracer of matter, such as galaxy lensing or CMB lensing. 
Since these galaxies produce some IR emission that contributes to the CIB, we decompose the unlensed CIB emission $T^0$ into a high-redshift component $T^{{\rm high}-z}$ (uncorrelated with the tracer), and a component $T_g$ originating from the tracers themselves, or objects correlated with them. 
Thus $T^0 = T_{{\rm high}-z} + T_g$. 
Since the $\hat \phi$ estimator is quadratic in $T$, $\hat \phi \sim T T$, the cross-correlation $\langle \hat \phi \delta_g \rangle$ contains a term of the form $\langle T_g T_g \delta_g \rangle \propto  B_{\delta_g T_g T_g}$, proportional to the hybrid bispectrum of two powers of $T_g$ and one power of $\delta_g$.

Following the Appendix of \cite{2017arXiv170506751F}, the self-lensing bias to the cross-correlation $C_L^{\hat{\phi} \times g}$ is given by
\beq
\left( \Delta  C_L^{\hat{\phi} \times g} \right)_{\rm SL} = N_\vL  \int d \chi \  \frac{W^g(\chi) (W^{\rm CIB}(\chi))^2}{\chi^2} \ \mathcal{B}(\k = \vL / \chi; \chi),
\label{eq:cross}
\eeq
with
\beq
\mathcal{B}(\k; \chi) \equiv \int \frac{d^2 \q}{(2\pi)^2} \  \frac{f_{\q \chi, \k \chi - \q \chi}}{2C^\text{total}_{\q \chi} C^\text{total}_{\k \chi - \q \chi}} \ B_{\delta T_g T_g}(\k, \q, -\k - \q; \chi) \,.
\label{eq:trianglePS}
\eeq
In the expression above, the momenta $\k, \q$ and $ -\k - \q$
lie on a plane perpendicular to the line of sight.

{\bf Auto-correlation}

Since the true lensing potential $\phi$ is also a tracer of the matter at low redshift, the previous discussion implies a bias to the auto-correlation of the reconstructed lensing potential:
\beq
\left( \Delta  C_L^{\hat{\phi} \hat{\phi}} \right)_{\rm SL} \approx 2 \left( \Delta  C_L^{\hat{\phi} \phi} \right)_{\rm SL} + {\rm secondary \ contractions}
\label{eq:auto}
\eeq
where the combinatorial factor of 2 arises from the two possible choices of $\hat \phi$ as a tracer. From Eq. \eqref{eq:cross}, we have
\beq
\left( \Delta  C_L^{\hat{\phi} \phi} \right)_{\rm SL} = N_\vL  \int d \chi \  \frac{W_L^{\phi_{\rm CIB}}(\chi) (W^{\rm CIB}(\chi))^2}{\chi^2} \ \mathcal{B}(\k = \vL / \chi; \chi),
\label{eq:crossphi}
\eeq
with $W_L^{\phi_{\rm CIB}}(\chi) = - 2 W^{\kappa_{\rm CIB}}(\chi)/L^2$ and $W^{\kappa_{\rm CIB}}$ as defined in Sec. \ref{sec:formalism}.

In addition to the term Eq.~\eqref{eq:crossphi}, secondary contractions appear in Eq.~\eqref{eq:auto}. To understand the origin of such terms, we once again separate the CIB emission into a high and low $z$ components.
Schematically, the auto correlation is given by $\langle \hat \phi \hat \phi \rangle \sim \langle (TT) (TT)\rangle$. Expanding the lensed $T$, we get `primary contractions' of the form $\langle (T_g T_g) (T_{{\rm high}-z} T_{{\rm high}-z}) \rangle$, which correspond to the ones in Eq. \eqref{eq:auto}, since $\langle  (T_{{\rm high}-z} T_{{\rm high}-z}) \rangle \sim \phi$. In addition, we also get `secondary contractions' of the form $\langle (T_g T_{{\rm high}-z}) (T_g T_{{\rm high}-z} ) \rangle$, which are the same order in the expansion as the primary contractions, but more involved to calculate, since the integrals do not separate as in Eqs. \eqref{eq:cross} and \eqref{eq:trianglePS}.
Similar contractions also appear in CMB lensing reconstruction (e.g., in the $N^{(1)}$ bias), and are highly suppressed compared to the primary contractions.  
However, this suppression is not guaranteed in our case. For example, secondary contractions of foreground contamination of CMB lensing has been estimated to be of the same order of magnitude as the primary ones \cite{2014JCAP...03..024O, 2017arXiv170506751F}.

\subsection{Mitigation strategies}

Controlling the biases presented above is crucial, in order to provide a convincing measurement of the lensing auto or cross-correlation.
In this section, we propose several mitigation approaches.

{\bf Cross Correlation} 

First, in cross-correlation with low-$z$ tracers,
the $N_L^{(0)}$ and $N_L^{(1)}$ terms only act as extra noise, but not as a bias.
In practice, this noise can usually be evaluated from the data itself.
As a result, the most concerning bias in cross-correlation is the self-lensing bispectrum.
We propose three methods to quantify this self-lensing bispectrum.

- \textit{Emission template}:
For a given tracer sample (e.g., galaxy catalog), we can build a template of the emission from the tracer sample at CIB frequencies $T_g$.
Because the bispectrum $B_{\delta T_g T_g}$ sourcing the self-lensing bias involves two powers of $T_g$,
having the correct galaxy flux fluctuations is important.
The optical or infrared brightness of the galaxies can inform us on their relative brightness at CIB frequency.
The overall normalization of the template map is obtained by cross-correlating the template with the CIB map.  
We can then apply the quadratic estimator to this template map and cross-correlate the result with the tracer map  to estimate the self-lensing bias.  
Alternatively, the template can be subtracted directly from the CIB map before applying the lensing quadratic estimator.

- \textit{Source hardening} :
Techniques similar to `source hardening' \cite{2014JCAP...03..024O} or the related `bias hardening' \cite{2013MNRAS.431..609N} used in CMB lensing are applicable to CIB lensing. 
If we assume a known shape for the source bispectrum $B_{\delta T_g T_g}$, an optimal estimator 
for its amplitude can be derived \cite{2007PhRvD..76d3510S, 2014JCAP...03..024O}. Given an amplitude for the source bispectrum, it can be propagated to the bias to CMB lensing by evaluating Eq. \eqref{eq:cross}. 
The source estimator and the lensing estimator are not orthogonal, but can be combined into unbiased estimates of the lensing potential and amplitude of sources, by inverting a response matrix as in \cite{2014JCAP...03..024O}.

- \textit{Separate shear and dilation estimation}: 
As discussed in the introduction (and \cite{2012PhRvD..85d3016B}), the quadratic estimator on large scales combines estimates of shear and dilation to reconstruct the lensing potential.
It is however possible to split it into a shear-only and dilation-only estimators \cite{2012PhRvD..85d3016B, 2017arXiv170902227P}. 
While the statistical power of the shear-only and dilation-only estimators is approximately equal, 
it is reasonable to assume that point source contamination, such as in self-lensing, should mostly affect the dilation estimator but not the shear estimator.
We leave investigating this conjecture for future work, and note that this could be a very effective technique to isolate contamination in both CIB and CMB lensing.

- \textit{Using disjoint sets of multipoles:}
In this case, as for self-lensing, lensing reconstruction and cross correlation using disjoint $\vl$ can help. For example, we could reconstruct $\hat{\phi}_{\vL}$ only using CIB multipoles $\ell > \ell_{\rm cut}$ and then perform the cross correlation only for $L <  \ell_{\rm cut}$. 
This ensures that the map used in the reconstruction $T(\ell > \ell_{\rm cut})$ is uncorrelated with the tracer map on the scales used for the cross-correlation.  
This eliminates the equilateral self-lensing bispectrum configuration,
and can thus limit the impact of contaminants that are common to the tracers and CIB map , e.g., residual Galactic dust.

 \textbf{Auto Correlation:}
 
Similar mitigation techniques apply for the auto-power spectrum;
we summarize these below.
In auto-correlation, $N^{(0)}$ and $N^{(1)}$ constitute a bias, contrary to the cross-correlation case.
These biases are usually computed with simulations and subtracted from the $\hat{\phi}$ power spectrum.  
However, this requires correctly simulating the intrinsic trispectrum of the CIB to much better precision than the lensing-induced trispectrum.  This can be challenging, since the intrinsic trispectrum is not well constrained, and even though it can be predicted in the halo model, achieving the required accuracy may not be straightforward. 
For this reason, auto-correlation analyses may be more challenging.
Most mitigation strategies presented for the cross-correlation also apply in auto-correlation. Here we focus on the differences.

- Making an \textit{emission template} from low-redshift galaxies is likely to be more challenging in auto-correlation than in cross: one would need a flux-weighted template of every source contributing to the low-$z$ part of the CIB, not only the sources from a given tracer catalog.  
To the extent that the emission traces matter and that the redshift-dependent IR luminosity function is known, a combination of galaxy surveys covering most of the CIB redshift distribution might provide the required template.

- \textit{Source hardening} seems a more promising avenue. In this case, we must assume a known shape for the intrinsic source bispectrum and trispectrum, and measure their amplitudes with appropriate estimators \cite{2014JCAP...03..024O}. This procedure is complicated by the presence of secondary contractions.

- \textit{Disjoint $\vl$ reconstruction} can be used in the auto-correlation as well. This method can be generalized and made more optimal by partitioning Fourier space into annuli, and evaluating the $\phi$ power spectrum only using the cross correlation between estimates reconstructed from modes belonging to different annuli \cite{2010arXiv1011.4510S}.

- Finally we note that part of the non-Gaussianity of the CIB is due to non-linear gravitational evolution. In particular, non-linear coupling between long and short wavelength modes produces a modulation in the small-scale power spectrum that is proportional to the long wavelength mode. The part of the trispectrum generated by this coupling can be undone by \textit{tidal reconstruction} \cite{2016arXiv161007062Z}.

\section{Bias from CIB lensing to CMB lensing reconstruction}
\label{sec:biasCMB}

In this section, we switch focus and consider CIB lensing not as the signal of interest, but as a contaminant to CMB lensing.
CMB lensing reconstruction typically operates on a cleaned temperature map.
Imperfect subtraction of the CIB (or any other foreground) biases the auto and cross-spectra of the reconstructed CMB lensing potential.
One factor that is often neglected is that some of the foregrounds lie at cosmological distance from us, and are therefore lensed.  
In this section, we present the extra bias to CMB lensing due to residual lensed CIB.  

Suppose that the cleaned CMB map $T$ contains not only the true lensed CMB $T_\text{CMB}$, but also some lensed CIB $T_\text{CIB}$.
This lensed CIB is either the full CIB emission, or the reduced CIB residual if some foreground reduction is applied.
 \beq
  T = T_\text{CMB} + T_\text{CIB} = \tc^0 - \mbf{\nabla} \phic \cdot \mbf{\nabla} \tc^0 + \ti^0 - \mbf{\nabla} \phii \cdot \mbf{\nabla} \ti^0 + \ldots
 \eeq
We carefully distinguish the CMB lensing potential $\phic$, which receives contributions from redshift zero to the surface of last scattering, from the CIB lensing potential $\phii$, which is only sourced by lower redshift structures.  
As shown in \cite{2014ApJ...786...13V, 2014JCAP...03..024O}, neglecting the $\ti^0$ contribution results in a biased estimation of $\phic$. 
As before, we treat the auto and cross-correlations separately.

 \textbf{Cross Correlation:}
 
 Suppose we ignore the presence of CIB contamination and apply the quadratic estimator to $T$, to get an estimate of $\hat{\phi}_{\rm CMB} = \mathcal{Q}_{\rm CMB}[T]$, where the subscript `CMB' on the quadratic estimator $\mathcal{Q}$ denotes that the weights used in the estimator are those appropriate to the CMB lensing reconstruction.  
 The cross-correlation with tracers $\delta_g$ then contains terms of the form $\langle \mathcal{Q}_{\rm CMB}[\ti] \ \delta_g \rangle$. 
 When replacing the lensed $\ti$ by the unlensed $\ti^0$, we get bispectrum terms $\langle \mathcal{Q}_{\rm CMB}[\ti^0] \ \delta_g \rangle \propto B_{T_g T_g \delta_g}$, which have been discussed in previous work.
But we also note that $\mathcal{Q}_{\rm CMB}[\ti] - \mathcal{Q}_{\rm CMB}[\ti^0]$ is effectively a reconstruction of $\phii$. This reconstruction is both suboptimal and biased because the CMB, rather than the CIB weights have been used\footnote{The optimal and unbiased reconstruction of $\phii$ would have been $\mathcal{Q}_{\rm CIB}[\ti]$.}. Nonetheless, if we write $\langle \mathcal{Q}_{\rm CMB}[\ti](\vL) \rangle = R(L) \phii(\vL)$ in terms of a response function $R$, then the cross correlation with tracers is further biased by a factor $R(L)\langle \phii \ \delta_g \rangle$.

\textbf{Auto Correlation:}

Similarly, the auto-correlation will be enhanced by terms of the form $R(L)\langle \phii \ \phic \rangle(L)$, $R^2(L)\langle \phii \ \phii \rangle(L)$, as well as the trispectrum  term $\mathscr{T}_L$ of the form of Eq.~\eqref{eq:T0}, but with the CIB lensing weights replaced with the CMB lensing ones.
Quantifying this effect is an interesting and important problem, but it is outside the scope of this paper.

\section{Conclusions}
\label{sec:conclusion}

Intensity mapping experiments such as 21cm, H$\alpha$, Lyman-$\alpha$, CO or CII \cite{2017arXiv170909066K} will probe large volumes of the universe.
Lensing reconstruction from these intensity maps has the potential to become a powerful cosmological tool,
by enabling measurements of the amplitude of structure at intermediate redshifts between galaxy surveys and the CMB.
This constitutes a new regime for lens reconstruction.
While the source field is continuous with small fluctuations (like in CMB lensing),
it is also generically non-Gaussian, and a full statistical prior is not readily available (like in galaxy lensing).
Furthermore, intensity maps may contain some redshift information, allowing the full 3D gravitational potential to be estimated.
Because of smooth foregrounds in frequency space
(e.g., due to Galactic dust in 21cm \cite{2016MNRAS.456.3142S} or continuum fitting in the Lyman-$\alpha$ forest),
wavevectors with low radial components may be unusable,
which complicates the lens reconstruction.

In this paper, we have addressed the issue of the non-Gaussianity of the source field.
We have generalized the quadratic estimator formalism commonly used in CMB lensing to any weakly non-Gaussian source field
by deriving the optimal weights.
We have computed the effect of non-Gaussianity on the lensing noise and shown that the relevant measure of non-Gaussianity is the ratio of the trispectrum to the squared power spectrum, divided by the number of modes.

Beyond the statistical error, the non-Gaussianity of the unlensed map also produces systematic biases in the lens reconstruction, both in auto and cross-correlation.
We have explored in detail the non-Gaussian noise biases, the effect of foregrounds, and `self-lensing' due to the lensing of high-redshift sources by the halos low-redshift sources.
We have proposed various means to reduce the non-Gaussianity of the source (e.g., by masking the rare bright point sources), to avoid it (e.g., using disjoint multipoles or shear/dilation estimators) and to subtract it (e.g., using emission templates or source/bias hardening).
Overall, these biases appear easier to mitigate in cross-correlation than in auto-correlation.

Since high quality CIB intensity maps already exist, we have applied our formalism to the case of CIB lensing.
We forecast that CIB lensing is detectable with current Planck data, and will be measured at high precision in future sub-millimeter experiments like CCAT-Prime.
Because the CIB is at lower redshift than most proposed intensity mapping surveys, the non-Gaussianity is expected to be larger.
Furthermore, the redshift distribution of the CIB is complex (dependent on the multipole considered), somewhat uncertain, and extremely wide, extending all the way to redshift zero.
This produces a larger self-lensing bias, compared to line intensity maps where the redshifts are known precisely.
CIB lensing is therefore a very useful stepping stone towards lensing reconstruction of intensity mapping experiments.

CIB lensing will also provide a measurement of the amplitude of structure at intermediate redshifts,
although its cosmological interpretation may at first be affected by uncertainties in the CIB redshift distribution.
Turning this around,
CIB lensing can be used to constrain the redshift distribution of the CIB,
by measuring distance ratios (sometimes referred to as `cosmography') between CIB lensing and CMB or galaxy lensing,
or by assuming a known cosmology.
In other words, measuring CIB lensing can provide useful insights on star formation history. 

Finally, because residual CIB will be present in cleaned CMB maps, the lensing of the CIB will be a bias in CMB lensing. 
We have described these biases, but leave their evaluation to future work.


\acknowledgments
We are grateful to Marcelo Alvarez, Emanuele Castorina, Anthony Challinor, William Coulton, Olivier Dor\'e, Simon Foreman, Chen He Heinrich, Daniel Lenz, Shirley Ho, Wayne Hu, Uro${\rm \check{s}}$ Seljak, Blake Sherwin, Kendrick Smith, Alex van Engelen, Martin White, Michael Wilson and Hong-Ming Zhu for useful conversations and comments.  
We thank the referee for useful comments.
This work used resources of the National Energy Research
Scientific Computing Center, a DOE Office of
Science User Facility supported by the Office of Science
of the U.S. Department of Energy under Contract No.
DE-AC02-05CH11231.
ES is supported by the National Science Foundation grant NSF AST1311756 and by the Chamberlain fellowship at Lawrence Berkeley National Laboratory.
SF thanks the Miller Institute for Basic Research in Science at the University of California, Berkeley for support.  
The Flatiron Institute is supported by the Simons Foundation.

\appendix

\section{CIB halo model}
\label{app:cib_halo_model}

\subsection{CIB monopole and power spectrum}

We follow the halo model implementation from \cite{2012A&A...537A.137P, 2014MNRAS.439..143P}.
Consider a galaxy with luminosity $L_\nu$, i.e. such that the power emitted in frequency band $d\nu$ is $L_\nu d\nu$, the observed flux per unit frequency is
\beq
S_\nu = \frac{a L_{\nu/a}}{4\pi \chi^2} \text{    [power/area/frequency]},
\eeq
where $a$ is the scale factor and $\chi$ the comoving distance. Note this equation does not involve the usual luminosity distance because we are not considering the bolometric flux, but a band-limited flux.
The observed CIB specific surface brightness $I_\nu$ is the sum of the fluxes from all galaxies within an observed solid angle $d\Omega$:
\beq
I_\nu = \int \frac{dN_\text{gal}}{d\Omega} \; S_\nu \text{    [power/area/solid angle/frequency]}.
\eeq
Introducing the comoving volume $dV = d\Omega \chi^2 d\chi$, and the galaxy number density contrast $\delta_\text{gal}(S_\nu)$, this can be expressed as
\beq
\delta I_\nu = \int d\chi \; \chi^2 \; 
\int dS_\nu\; \frac{d\bar{N}_\text{gal}}{dS_\nu dV} \; 
S_\nu \; \delta_\text{gal}(S_\nu).
\eeq
The flux distribution functions of galaxies $d\bar{N}_\text{gal}/dS_\nu dV$ for the Planck frequency bands are taken from the `two star-forming modes' (2SFM) model in \cite{2012ApJ...757L..23B}. We apply the flux cuts relevant for Planck (see table 1 in \cite{2014A&A...571A..30P}), and for CCAT-Prime.
The model we implement makes the simplifying but unphysical assumption that all galaxies cluster identically, regardless of their luminosity. With this assumption, the equation above becomes:
\beq
\delta I_\nu = \int d\chi\; \chi^2\;
\bar{j}_\nu
\delta_\text{gal},
\eeq
where we have introduced:
\beq
\bar{j}_\nu = \int dS_\nu\; \frac{d\bar{N}_\text{gal}}{dS_\nu dV} \; 
S_\nu.
\eeq
Note that our definition differs by a factor $a/\chi^2$ from other sources in the literature.

Hence the CIB intensity power spectrum:
\beq
C_\ell^{I_\nu I_{\nu^\prime}}
=
\int d\chi \; \chi^2 
\bar{j}_\nu \bar{j}_{\nu^\prime} \;
P_\text{gal}^{3d} \left( k = \frac{\ell+1/2}{\chi}, z \right).
\eeq
The power spectrum of galaxies is then evaluated from a halo model, which produces a 2h-term, 1h-term (or halo shot noise) and a 1-galaxy term (or galaxy shot noise). 
The clustering of galaxies (2-halo and 1-halo terms) is computed using a halo occupation distribution (HOD) from \cite{2010ApJ...719...88T}:
\beq
\left\{
\bal
P_\text{gal}^{1h}(k) &= I_{02}(k)\\
P_\text{gal}^{2h}(k) &= I_{11}^2(k) P^\text{lin} (k),\\
\eal
\right.
\eeq
where:
\beq
I_{ij}(k_1, ..., k_j) \equiv 
\int dm \frac{dn}{dm} 
b_i
\left( \frac{N_\text{gal}(m)}{\bar{n}_\text{gal}} \right)^j
u(k_1)...u(k_j).
\eeq
Here $N_\text{gal}(m)$ and $\bar{n}_\text{gal}$ define the HOD, and $u(k)$ is the normalized Fourier transform of the halo profile \cite{2010ApJ...719...88T}.
Following \cite{2014MNRAS.439..143P}, we vary the HOD depending on the frequency band observed.
The shot noise power spectrum is computed as
\beq
d C_\ell^\text{gal. shot} = \frac{dN}{d\Omega} S_\nu
\text{   , so    }
C_\ell^\text{gal. shot} = 
\int d\chi \; \chi^2 \;
\underbrace{
\int dS_\nu\; \frac{d\bar{N}_\text{gal}}{dS_\nu dV} \; S_\nu^2}
_{\equiv \bar{j}_{\nu, 2}}.
\eeq
Finally, our model for the CIB power spectrum becomes:
\beq
C_\ell^{I_\nu I_{\nu^\prime}}
=
\int d\chi \; \chi^2 
\left[
\bar{j}_\nu \bar{j}_{\nu^\prime}
P_\text{gal}^\text{1h+2h}
+ 
\bar{j}_{\nu, 2}
\right].
\eeq
Several limitations should be noted.
First, the CIB emission is assumed independent of galaxy mass.
Second, for this approach to be consistent, there has to exist a mass-flux relation such that the HOD in \cite{2010ApJ...719...88T} produces the flux counts of \cite{2012A&A...537A.137P}, which is not guaranteed.
Finally, for the mass integrals to converge, one needs to integrate to very low halo masses, $\simeq 10^{10} M_\odot$, where the HOD may be uncertain.

\subsection{CIB trispectrum}
\label{app:trispectrum_hm}

The power spectrum of the reconstructed lensing potential is a four point function of the temperature.
Therefore, the unlensed CIB trispectrum contributes to the lensing power spectrum.
Similarly to the case of the power spectrum, the halo model relates the CIB trispectrum to the galaxy trispectrum, and eventually to the matter power spectrum, bispectrum and trispectrum:
\beq
\mathcal{T}_{\vl_1, \vl_2, \vl_3, \vl_4}^\text{CIB}
=
\int d\chi \; \chi^2 
\left[
\bar{j}_\nu^4\;
\mathcal{T}_\text{gal}^\text{1h+2h+3h+4h}
+ 
\bar{j}_{\nu, 4}
\right],
\label{eq:trispectrum_halo_model}
\eeq
with
\beq
\bar{j}_{\nu, 4} = \int dS_\nu\; \frac{d\bar{N}_\text{gal}}{dS_\nu dV} \; S_\nu^4.
\eeq
As shown in Fig.~\ref{fig:cib_trispectrum}, our halo model trispectrum is very sensitive to the minimum redshift (or distance) in the integral above.
In order to evaluate Eq~\eqref{eq:trispectrum_halo_model}, we need to compute
the galaxy trispectrum $\mathcal{T}_\text{gal}$ in the halo model \cite{2013PhRvD..87l3504T}:
\beq
\left\{
\bal
\mathcal{T}_\text{gal}^\text{1h}(\k_1, \k_2, \k_3, \k_4) &= I_{04}(k_1, k_2, k_3, k_4) \\
\mathcal{T}_\text{gal}^\text{2h}(\k_1, \k_2, \k_3, \k_4) &=  P_m(k_{12}) I_{12}(k_1, k_2) I_{12}(k_3, k_4) + 2 \text{  perm.} \\
                                                                   &+ P_m(k_1) I_{11}(k_1) I_{13}(k_2, k_3, k_4) + 3 \text{  perm.}\\
\mathcal{T}_\text{gal}^\text{3h}(\k_1, \k_2, \k_3, \k_4) &= B_m(\k_1, \k_2, \k_{34}) I_{11}(k_1) I_{11}(k_2) I_{12}(k_3, k_4) + 5 \text{  perm.}\\
\mathcal{T}_\text{gal}^\text{4h}(\k_1, \k_2, \k_3, \k_4) &= T_m(\k_1, \k_2, \k_3, \k_4) I_{11}(k_1) I_{11}(k_2) I_{11}(k_3) I_{11}(k_4),\\
\eal
\right.
\eeq
where $P_m$, $B_m$ and $T_m$ are the matter power spectrum, bispectrum and trispectrum, respectively.

When applying the quadratic estimator, the low lensing multipoles $L$ are best reconstructed, and this reconstruction relies mostly on the high map multipoles $\ell$. 
The trispectrum configuration of interest is therefore `almost-collapsed', $\mathcal{T}^\text{CIB}_{\ell, L-\ell, \ell ', -L - \ell '}$, with $L \ll \ell$.
This is convenient for two reasons. 
First, the almost-collapsed trispectrum can be measured from the map without having to implement a general trispectrum estimator (see App.~\ref{app:trispectrum_measurement}).
Second, in this limit, the halo model expressions simplify.
Indeed, for the galaxy number density or any general 3D field, there exists a consistency relation between the almost-collapsed trispectrum, the exactly-collapsed trispectrum $\mathcal{T}^\text{gal}(\k, -\k, \k', - \k')$, and the response of the power spectrum to a matter overdensity:
\beq
\underbrace{\mathcal{T}^\text{gal}(\k, \K-\k, \k', -\K - \k')}
_{\text{almost-collapsed}}
\simeq
\underbrace{\mathcal{T}^\text{gal}(\k,-\k, \k', - \k')}
_{\text{exactly-collapsed}}
+
\underbrace{\frac{\partial P^\text{gal}(k)}{\partial \delta_m}
\frac{\partial P^\text{gal}(k)}{\partial \delta_m}
P^\text{lin}(K)}
_{\text{power spectrum response}}.
\eeq
This relation was derived in the context of the covariance of measured power spectra, and the second term on the right hand side is then interpreted as the supersample variance \cite{2013PhRvD..87l3504T, 2014JCAP...05..048C, 2014PhRvD..90l3523S}.
In \cite{2013PhRvD..87l3504T}, it was shown that the halo model satisfies this consistency relation.
This is convenient, because it means that we only need to evaluate the halo model trispectrum in the exactly collapsed limit.

We follow \cite{2013PhRvD..87l3504T} for the expressions of the halo model trispectrum.
The shot noise and 1-halo trispectra are easy to evaluate, depending only on the mass function, HOD and halo profile.
The 2-halo further depends on the matter power spectrum, and is also straightforward to compute.
In the exactly-collapsed limit, the 3-halo term $\mathcal{T}^{3h}$ cancels because the matter bispectrum is zero in the collapsed configuration.
The 4-halo trispectrum is related to the matter trispectrum. 

For the purpose of lens reconstruction, we focus on the collapsed trispectrum 
$ \langle \mathcal{T}^\text{gal}(\k,-\k, \k', - \k') \rangle$, averaged over the angle between $\k$ and $\k'$, and with $k'=k$.
Indeed, we mostly reconstruct the low $L$ lensing modes, and from only high $\ell$ temperature modes, which justifies the collapsed limit.
Furthermore, as can be seen from Fig.~\ref{fig:lensing_contribution}, the range of multipoles which contribute to the (Gaussian) signal-to-noise is quite narrow, which justifies considering $k'=k$. Finally, the power spectrum of the reconstructed lensing at fixed $\vL$ is effectively an average over the angle between $\vl$ and $\vl^\prime$.
It is true that this average involves some angular weights, but this is not expected to change the result by more than a factor of a few, which is sufficient here.
In this very specific limit, the matter trispectrum takes a simple form   $\langle \mathcal{T}_\text{m}(\k, -\k, \k', -\k') \rangle \simeq (232/441) P_m(k)^3$ \cite{1999ApJ...527....1S},
and the halo model terms become:
\beq
\left\{
\bal
\langle\mathcal{T}_\text{gal}^\text{1h}(\k, -\k, \k', -\k')\rangle &= I_{04}(k, k, k, k) \\
\langle\mathcal{T}_\text{gal}^\text{2h}(\k, -\k, \k', -\k')\rangle &= 4 P_m(k) I_{11}(k) I_{13}(k,k,k) + 2 P_m(k) I_{12}(k,k)^2 \\
\langle\mathcal{T}_\text{gal}^\text{3h}(\k, -\k, \k', -\k')\rangle &= 0\\
\langle\mathcal{T}_\text{gal}^\text{4h}(\k, -\k, \k', -\k')\rangle &= \frac{232}{441} P_m(k)^3I_{11}(k)^4 .\\
\eal
\right.
\eeq
Finally, we follow \cite{2014JCAP...05..048C} for the expression of the power spectrum response $\frac{\partial P^\text{gal}(k)}{\partial \delta_m}$ in the halo model.

\section{Deriving the lensing kernel for the CIB lensing quadratic estimator}
\label{app:cib_lensing_kernel}

In this appendix, we derive the lensing efficiency kernel for the CIB lensing quadratic estimator. 
In other words, we wish to derive the redshift range probed by the CIB lensing estimator. 

In the case of CMB lensing, this question is trivial, since the CMB is sourced at a single redshift of $1100$.
In the case of galaxy lensing, there is a well defined number density of galaxies $dn/dz$. 
The shear estimator is linear in the individual galaxy shapes.
If the signal-to-noise on each galaxy shape was the same, regardless of galaxy redshift, then the observed shear is simply the average of the shears defined at each galaxy redshift, weighted by $dn/dz$.

In the case of CIB lensing, this question is less straightforward.
 Indeed, depending on the multipole, the unlensed CIB is sourced by different redshift ranges: the 2-halo, 1-halo and 1-galaxy terms in the CIB power spectrum have different redshift distributions (see Fig. 12 in \cite{2013A&A...557A..66B}).
The quadratic estimator mixes the CIB multipoles to reconstruct the lensing potential, in a way that depends on the multipole of lensing to be reconstructed. Thus, different multipoles of the CIB lensing quadratic estimator probe different redshifts.

We decompose the unlensed CIB emissions into statistically independent redshift slices:
\beq
T^0_\vl = \int dz \; T^z_\vl,
\text{  with }
\langle T^z_\vl T^{z^\prime}_{\vl^\prime} \rangle \propto \delta^D_{z-z^\prime}.
\eeq
As a result, the unlensed CIB power spectrum can be decomposed as
\beq
C^0_\ell = \int dz \frac{d C^0_\ell}{d z},
\text{  where }
\langle T^z_\vl T^{z^\prime}_{\vl^\prime} \rangle = \delta^D_{z-z^\prime} (2\pi)^2 \delta^D_{\vl+\vl^\prime} \frac{d C^0_\ell}{d z}
\eeq

We then define $\phi^z$ to be the lensing potential for a source at redshift $z$.
The unlensed CIB emission at redshift $z$ is lensed by the potential $\phi^z$. 
Thus,
\beq
\langle T^z_{\vl} T^z_{\vL-\vl} \rangle
=
\phi^z_\vL
\;
f^z_{\vl, \vL-\vl}
+
\mathcal{O}(\phi^2 C^0) ,
\eeq
where
\beq
f^z_{\vl, \vL-\vl} 
=
-
\vL \cdot
\left[ 
\vl \frac{dC^0_{\ell}}{dz}
+
 (\vL - \vl) \frac{dC^0_{\vL - \vl}}{dz}
 \right] .
\eeq
In particular,
\beq
f_{\vl, \vL-\vl} 
=
\int dz f^z_{\vl, \vL-\vl} .
\eeq

As a result, the basic quadratic estimator measures
\beq
\langle \hat{\phi}_{\vL, \vl} \rangle
=
\int dz \;
\phi_\vL^z \;
\frac{f^z_{\vl, \vL-\vl}}{f_{\vl, \vL-\vl}}.
\eeq
We can finally express the quadratic estimator as a weighted sum of the lensing potential over the various redshifts slices, with the weights
\beq
\langle \hat{\phi}_\vL \rangle
\propto
\int dz\;
\phi_\vL^z \;
\int \frac{d^2\vl}{(2\pi)^2}
\frac{f^z_{\vl, \vL-\vl}}{f_{\vl, \vL-\vl}}
\frac{1}{\sigma^2_{\vL, \vl}}.
\eeq
In conclusion, the measured CIB lensing convergence can be written as
\beq
\kappa_{\text{CIB}\; \vL} =  \int d\chi \; W^{\kappa_\text{CIB}}(\chi, \vL) \; \delta_m(\vec{k} = \vL/\chi, \chi),
\eeq
where,
\beq
W^{\kappa_\text{CIB}} (\chi, \vL) = \; \int d\chi_S
W^{\text{CIB}} (\chi_S, \vL) \; 
 W^\kappa(\chi, \chi_S),
\eeq
\beq
W^\kappa (\chi, \chi_S) = \frac{3}{2} \left( \frac{H_0}{c} \right)^2 \Omega_m^0 \frac{\chi}{a(\chi)} \left( 1 - \chi/\chi_S \right),
\eeq
and
\beq
W^\text{CIB} (\chi_S, \vL)
\propto
\frac{dz}{d\chi_s} \int \frac{d^2\vl}{(2\pi)^2}\;
\frac{f^z_{\vl, \vL-\vl}}{f_{\vl, \vL-\vl}}
\frac{1}{\sigma^2_{\vL, \vl}},
\text{   normalized by }
\int d\chi_S  W^\text{CIB} (\chi_S, \vL) = 1.
\eeq

A few comments are in order.
First, the lensing kernel $W^{\kappa_\text{CIB}} (\chi, \vL)$ depends on the multipole $\vL$ considered. This is not the case in CMB lensing, due to fact that the unlensed CMB comes from a single redshift. This is also different from the galaxy lensing case.

Second, the lensing kernel for the CIB quadratic estimator is determined by the redshift distribution of the unlensed CIB power spectrum 
$\frac{dC^0_\ell}{dz}$ as well as the noise power spectrum of the CIB map. 
Intuitively, the minimum variance quadratic estimator weighs the various CIB  multipoles differently depending on their signal-to-noise. Since different redshifts contribute differently to the various CIB multipoles, this weighting will up-weight or down-weight  certain redshift ranges.

\section{Non-Gaussianity and power spectrum signal-to-noise}
\label{app:powerspectrum_snr_ng}

Before looking at the case of lensing, as a first step we compute the angular power spectrum covariance in both the Gaussian and Poisson cases.
For multipoles above $\ell \sim 1000$, the CIB is dominated by the 1-halo or 1-galaxy terms. In this regime, the statistics of the field are close to a uniform Poissonian sampling, rather than a Gaussian random field. 
Below, we compute and interpret intuitively the reduction in signal-to-noise due to the Poisson sampling.
This intuition is useful to understand the effect of non-Gaussianity on the lensing signal-to-noise, as described in the next section.

\subsection{Gaussian random field}

For a Gaussian random field, we know that the various $\ell$-modes are independent. This leads to the well-known formula for the power spectrum covariance:
\beq
\text{cov}\left[ C_\ell, C_{\ell^\prime} \right]
=
\delta^K_{\ell, \ell^\prime}
\frac{2 C_\ell^2}{N_{\text{modes } \ell}},
\eeq
where $N_{\text{modes } \ell} \approx 2f_{\rm sky} \ell \Delta \ell $  is the number of observed Fourier modes in a bin centered at $\ell$ and of width $\Delta \ell$.
As a result, the signal-to-noise ratio for the power spectrum is
\beq
\text{SNR}_\text{Gaussian}(C_\ell) = \sqrt{ \frac{N_{\text{modes } \ell}}{2} }.
\eeq
Intuitively, each mode contributes a signal-to-noise of $1/\sqrt{2}$, and the modes are all independent, so $\text{SNR} \propto \sqrt{ N_{\text{modes }\ell}}$.
As expected, the signal-to-noise becomes infinite in the limit of infinite number of modes: the only limit to the signal-to-noise is the number of modes.

\subsection{Non-Gaussian random field}

For a non-Gaussian random field, the situation is different. The various multipoles are no longer independent, which translates into correlations across power spectrum bins:
\beq
\text{cov}\left[ C_\ell, C_{\ell^\prime} \right]
=
\delta^K_{\ell, \ell^\prime}
\frac{2 C_\ell^2}{N_{\text{modes } \ell}}
+
\frac{\mathcal{T}_{\ell, -\ell, \ell^\prime, -\ell^\prime}}{V},
\eeq
where $V$ is the 3D volume or 2D area of the survey.
The trispectrum of the field thus adds a non-diagonal covariance term.
As a result, the signal-to-noise is reduced compared to the Gaussian case.
We explore this in more detail in the special case of the uniform Poisson sampling in the following.

\subsection{Uniform Poisson sampling}

A uniform Poisson sampling well describes a collection of unclustered point sources.
In the case, the power spectrum and trispectrum take simple forms that are independent of multipole:
\beq
\left\{
\bal
& C_\ell = \frac{\langle s^2 \rangle}{\bar{n}},\\
&\mathcal{T}_{\vl, -\vl, \vl^\prime, -\vl^\prime} = \frac{\langle s^4 \rangle}{\bar{n}^3},\\
\eal
\right.
\eeq
where each object has a flux $s$ and the mean number density of objects is $\bar{n}$.

In this simple case, the power spectrum is entirely determined by these two numbers. 
The most natural way of extracting this information would be to estimate $\bar{n}$ and $\langle s^2 \rangle$ directly, without computing the power spectrum.
One could simply count the number of objects. 
Intuitively, the signal-to-noise on this measurement should scale as $\sqrt{N_\text{gal}} = \sqrt{\bar{n} V}$. 
One would then estimate $\langle s^2 \rangle$ directly from the galaxy catalog. The signal-to-noise on the power spectrum will be determined by the amount of object-to-object fluctuations in the flux, i.e. it will scale as $\langle s^2 \rangle / \sqrt{\langle s^4 \rangle}$.

While measuring the power spectrum is a good way to extract the information for the Gaussian case, it is a clumsy approach for the Poisson case.
First, we can see that the various $C_\ell$ bins are correlated due to the trispectrum term. 
However, they are not $100\%$ correlated, so extracting all the information requires measuring all the available multipoles.
Indeed, the non-Gaussian formula for the signal-to-noise simplifies to
\beq
\bal
\text{SNR}_{\rm Poisson}(C_\ell) 
&=
\sqrt{\frac{N_{\text{modes} \ \ell}}{2}}
\frac{1}{\sqrt{1 + \frac{N_{\text{modes} \ \ell} \mathcal{T} }{2C^2 V} }} \\
&=
\sqrt{\frac{N_{\text{modes} \ \ell}}{2}}
\frac{1}{\sqrt{1 + \frac{N_{\text{modes} \ \ell} }{2 N_\text{gal}} \frac{\langle s^4 \rangle}{\langle s^2 \rangle^2} }}
.
\eal
\eeq
In particular, in the limit of infinite number of modes, the signal-to-noise is not infinite, but asymptotes to
\beq
\text{SNR}_\text{Poisson} (C_\ell, N_{\text{modes} \ \ell}\rightarrow \infty) = \sqrt{ N_\text{gal} }  \frac{\langle s^2 \rangle}{\sqrt{\langle s^4 \rangle}}.
\eeq
We thus obtain the expression expected for the information content of a Poisson sampling.

\section{Non-Gaussianity and lensing signal-to-noise}
\label{app:N0_nongausssource}

We go a step further and compute the total noise on the lens reconstruction when the unlensed source image is a uniform Poisson sampling.
In this simple limiting case, the power spectrum and trispectrum are independent of multipole.
This allows to evaluate Eq.~\eqref{eq:N0_nongausssource_nongaussweights} and~\eqref{eq:N0_nongausssource_gaussweights},
and gain some intuition for the magnitude of the non-Gaussian lensing noise.

\subsection{Non-Gaussian weights}

To evaluate Eq.~\eqref{eq:N0_nongausssource_nongaussweights}, we need to be careful about double counting modes. Noting that  $\hat{\phi}_{\vL, \vl} = \hat{\phi}_{\vL, \vL-\vl}$, we can replace the covariance matrix by
\beq
\left(\Sigma_{\vL}\right)_{\vl, \vl^\prime}
=
\frac{1}
{f_{\vl, \vL-\vl} \; f_{\vl^\prime, \vL-\vl^\prime}}
\left[
2 (2\pi)^2 \delta^D_{\vl^\prime-\vl} 
C^\text{total}_\vl C^\text{total}_{\vL-\vl}
+\mathcal{T}^0_{\vl, -\vl^\prime, \vL-\vl, \vl^\prime-\vL}
\right].
\eeq
If power spectrum and trispectrum are independent of $\ell$, and $C^\text{total}\simeq C^0$, the Sherman-Morrison formula allows to invert the covariance matrix as
\beq
\left(\Sigma_{\vL}\right)_{\vl, \vl^\prime}^{-1}
=
\frac{f_{\vl, \vL-\vl} \; f_{\vl^\prime, \vL-\vl^\prime}}
{2 \left(C^0\right)^2}
\left[
(2\pi)^2 \delta^D_{\vl^\prime-\vl} 
-
\frac{\mathcal{T}^0}
{2\left(C^0\right)^2 + \mathcal{T}^0 \left( \int\frac{d^2\vl}{(2\pi)^2} \right)}
\right],
\eeq
which we then integrate:
\beq
\int \frac{d^2\vl d^2\vl^\prime}{(2\pi)^4} \left(\Sigma_\vL\right)^{-1}_{\vl, \vl^\prime}
=
L^4
\frac{\left(\int \frac{d^2\vl}{(2\pi)^2}\right) \left(C^0\right)^2}
{2\left(C^0\right)^2 + \left(\int \frac{d^2\vl}{(2\pi)^2}\right) \mathcal{T}^0}.
\eeq
Therefore, the reconstruction noise on $\phi$ (Eq.~\eqref{eq:N0_nongausssource_nongaussweights}) becomes
\beq
N_\vL
=
\left[
\int \frac{d^2\vl d^2\vl^\prime}{(2\pi)^4} \; \left( \Sigma_{\vL} \right)^{-1}_{\vl, \vl^\prime}
\right]^{-1} \\
=
\underbrace{\frac{2}{L^4 \left(\int \frac{d^2\vl}{(2\pi)^2}\right)}}
_{\text{Gaussian } N_L}
+
\underbrace{\frac{\mathcal{T}^0}{L^4 (C^0)^2}}
_{\text{non-Gaussian correction}},
\eeq
or in terms of $\kappa$:
\beq
\bal
N_\vL^\kappa
&=
\underbrace{\frac{2\pi}{N_\text{modes}}}
_{\text{Gaussian noise}}
+
\underbrace{\frac{\mathcal{T}^0}{4 (C^0)^2}}
_{\text{non-Gaussian correction}},
\eal
\label{eq:poisson_ng_lensing_noise}
\eeq
where $N_\text{modes} \equiv 4\pi \int \frac{d^2\vl}{(2\pi)^2} =  \ell_\text{max}^2 - \ell_\text{min}^2$ and $\ell_\text{min}$ and $\ell_\text{max}$ define the range of observed multipoles used for the lens reconstruction.
The first term recovers the noise relevant for a Gaussian (instead of Poisson) white noise. This term goes to zero as the number of observed multipoles increases: if the modes are Gaussian, each of them adds independent lensing information, and the number of modes is the only limit to the reconstruction precision. 
The second term is due to the non-Gaussian nature of the unlensed map. Interestingly, it is independent of the number of observed multipoles.

In the limit of infinite number of observed multipoles, the reconstruction noise thus asymptotes to 
$N_\vL \rightarrow \frac{\mathcal{T}^0}{L^4 C^2}$.
This is the noise on the reconstructed lensing potential $\phi$. The corresponding noise on the lensing convergence $\kappa$ is 
\beq
N_\vL^\kappa
=
\frac{\mathcal{T}^0}{4 C^2}
=
\frac{1}{4}
\frac{\langle s^4 \rangle}{\langle s^2 \rangle^2}
\frac{1}{\bar{n}}.
\eeq
This shows that the noise on the reconstructed $\hat{\phi}_\vL$ is independent of the multipole $\vL$. This is the same $L$-dependence as the shape noise of galaxy shear.
As stated above, because the unlensed power spectrum is white ($C_\ell$ is independent of $\ell$), shear has no effect on the map and all the lensing information is reconstructed from magnification.
Hence, this noise does not originate from fluctuations in the shapes of galaxies, but rather in fluctuations in the local number density and brightness of galaxies. Locally, these fluctuations are degenerate with a local magnification and are therefore a source of noise for the lens reconstruction.

A more intuitive derivation is the following. 
Schematically, the observed surface brightness map can be described as $n s$, where $n$ is the number count of galaxies and $s$ is the galaxy brightness. Lensing conserves the surface brightness. Indeed, if the convergence is non-zero, the observed number counts are diluted as $\bar{n} \rightarrow \bar{n}/(1+2\kappa)$, but the individual galaxy brightnesses are enhanced as $\langle s \rangle \rightarrow \langle s \rangle (1+2\kappa)$, thus producing no net effect. On the other hand, the power spectrum of the map, which scales as $C^0_\ell = \langle s^2 \rangle / \bar{n} $, is modified as $C_\ell = C^0_\ell (1+2\kappa)$. As a result, the reconstruction noise on the convergence is directly related to the intrinsic fluctuations in the power spectrum:
\beq
\sigma_\kappa^2 = \frac{1}{4} \frac{1}{\text{SNR}_\text{Poisson}},
\eeq
where $\text{SNR}_\text{Poisson} = \frac{\mathcal{T}^0}{C^2}$ is the signal-to-noise ratio on the measured power spectrum, computed above. 
Finally, since this noise on the reconstruction of the local $\kappa$ is uncorrelated between positions, we get
\beq
N_\vL^\kappa = \sigma_\kappa^2 = \frac{1}{4} \frac{1}{\text{SNR}_\text{Poisson}} = \frac{\mathcal{T}^0}{4C^2},
\eeq
consistent with the result above.

\subsection{Gaussian weights}

When the power spectrum and trispectrum are independent of multipole, Eq.~\eqref{eq:N0_nongausssource_gaussweights} for the lens reconstruction noise takes the simple form:
\beq
\bal
N_\vL^\kappa
&=
\frac{2\pi}{N_\text{modes}}
+
\frac{\mathcal{T}^0}{4 (C^0)^2},
\eal
\eeq
Notice that the reconstruction noise in this case (Gaussian weights on a non-Gaussian unlensed map) is the same as when using the more optimal non-Gaussian weights. 
This coincidence occurs for the following reason: since the power spectrum and trispectrum are independent of multipole, both the non-Gaussian and Gaussian weights give equal weight to every temperature mode, resulting in the same estimator.
This is not the case for a general power spectrum and trispectrum.

\section{Measuring the trispectrum in the collapsed limit}
\label{app:trispectrum_measurement}

Here we present the estimator for the CIB trispectrum $\mathcal{T}^0_{\vl, \vL-\vl, \vl, -\vL-\vl}$  in the collapsed limit $L \ll \ell$ that we have used to derive the upper limits in Fig.~\ref{fig:cib_trispectrum}.
Let us define a filtered map $T_f(\vl)= W_f(\ell) T_\vl$, with the filter normalizes such that
\beq
\int \frac{d^2\vl }{(2\pi)^2} W_f^2(\ell) = 1.
\eeq
Let us further define the field $K(\x)$ in real space as $K(\n) = T_f^2(\n)$.  The power spectrum of $K$ is then
\begin{eqnarray}
\langle K(\vL) K(\vL') \rangle & = & \int \frac{d^2\vl d^2\vl^\prime}{(2\pi)^4} W_f(\ell) W_f(|\vL-\vl|) W_f(\ell') W_f(|\vL'-\vl'|) \langle T_\vl T_{\vL-\vl} T_{\vl'} T_{\vL'-\vl'} \rangle \nonumber \\
& = & (2\pi)^2 \delta^D_{\vL+\vL'} \times \left( 2 \int \frac{d^2\vl }{(2\pi)^2} W_f^2(\ell) W_f^2(|\vL-\vl|) C_\ell^{\rm total} C_{|\vL-\vl|}^{\rm total} \right. \nonumber
\\ &+& \left. \int \frac{d^2\vl d^2\vl^\prime}{(2\pi)^4} W_f(\ell) 
W_f(|\vL-\vl|) W_f(\ell') 
W_f(|\vL'-\vl'|) \mathcal{T}^0_{\vl, \vL-\vl, \vl', \vL'-\vl'} \right) .
\end{eqnarray}
In this limit, and for a narrow filter centered around $\ell_*$, we thus obtain:
\beq
C_L^{KK} = 
\left[
2
\underbrace{ \left( \int \frac{d^2\vl }{(2\pi)^2} W_f^4(\ell) \right)
(C_{\ell_*}^{\rm total})^2 }_\text{Gaussian part}
+
 \underbrace{ \langle \mathcal{T}^0_{\vl_*, \vL-\vl_*, \vl_*, -\vL-\vl_*} \rangle }_\text{angle-averaged trispectrum}
\right].
\eeq
To estimate the trispectrum, we subtract the Gaussian part of $C_L^{KK}$ measured on Gaussian simulations with the same power spectrum as the CIB.



\begin{thebibliography}{3}
\bibitem[Bartelmann \& Schneider(2001)]{2001PhR...340..291B} Bartelmann, M., \& Schneider, P.\ 2001, \physrep, 340, 291 
\bibitem[Schneider(2005)]{2005astro.ph..9252S} Schneider, P.\ 2005, arXiv:astro-ph/0509252 
\bibitem[Kilbinger(2015)]{2015RPPh...78h6901K} Kilbinger, M.\ 2015, Reports on Progress in Physics, 78, 086901 
\bibitem[Hildebrandt et al.(2017)]{2017MNRAS.465.1454H} Hildebrandt, H., Viola, M., Heymans, C., et al.\ 2017, \mnras, 465, 1454 
\bibitem[Becker et al.(2016)]{2016PhRvD..94b2002B} Becker, M.~R., Troxel, M.~A., MacCrann, N., et al.\ 2016, \prd, 94, 022002 
\bibitem[Mandelbaum et al.(2017)]{2017arXiv170506745M} Mandelbaum, R., Miyatake, H., Hamana, T., et al.\ 2017, arXiv:1705.06745 

\bibitem[Lewis \& Challinor(2006)]{2006PhR...429....1L} Lewis, A., \& Challinor, A.\ 2006, \physrep, 429, 1 
\bibitem[Hanson et al.(2010)]{2010GReGr..42.2197H} Hanson, D., Challinor, A., \& Lewis, A.\ 2010, General Relativity and Gravitation, 42, 2197 
\bibitem[Hu \& Okamoto(2002)]{2002ApJ...574..566H} Hu, W., \& Okamoto, T.\ 2002, \apj, 574, 566 
\bibitem[Hirata \& Seljak(2003)]{2003PhRvD..68h3002H} Hirata, C.~M., \& Seljak, U.\ 2003, \prd, 68, 083002 
\bibitem[Carron \& Lewis(2017)]{2017arXiv170408230C} Carron, J., \& Lewis, A.\ 2017, arXiv:1704.08230 
\bibitem[Sherwin et al.(2016)]{2016arXiv161109753S} Sherwin, B.~D., van Engelen, A., Sehgal, N., et al.\ 2016, arXiv:1611.09753 
\bibitem[Omori et al.(2017)]{2017arXiv170500743O} Omori, Y., Chown, R., Simard, G., et al.\ 2017, arXiv:1705.00743 
\bibitem[Planck Collaboration et al.(2016)]{2016A&A...594A..15P} Planck Collaboration, Ade, P.~A.~R., Aghanim, N., et al.\ 2016, \aap, 594, A15 
\bibitem[Ade et al.(2014)]{2014PhRvL.113b1301A} Ade, P.~A.~R., Akiba, Y., Anthony, A.~E., et al.\ 2014, Physical Review Letters, 113, 021301 
\bibitem[Bucher et al.(2012)]{2012PhRvD..85d3016B} Bucher, M., Carvalho, C.~S., Moodley, K., \& Remazeilles, M.\ 2012, \prd, 85, 043016 

\bibitem[Pratten \& Lewis(2016)]{2016JCAP...08..047P} Pratten, G., \& Lewis, A.\ 2016, \jcap, 8, 047 
\bibitem[Fabbian et al.(2017)]{2017arXiv170203317F} Fabbian, G., Calabrese, M., \& Carbone, C.\ 2017, arXiv:1702.03317 
\bibitem[Petri et al.(2017)]{2017PhRvD..95l3503P} Petri, A., Haiman, Z., \& May, M.\ 2017, \prd, 95, 123503

\bibitem[Puget et al.(1996)]{1996A&A...308L...5P} Puget, J.-L., Abergel, A., Bernard, J.-P., et al.\ 1996, \aap, 308, L5 
\bibitem[Planck Collaboration et al.(2014)]{2014A&A...571A..18P} Planck Collaboration, Ade, P.~A.~R., Aghanim, N., et al.\ 2014, \aap, 571, A18 
\bibitem[Lacasa et al.(2014)]{2014MNRAS.439..123L} Lacasa, F., P{\'e}nin, A., \& Aghanim, N.\ 2014, \mnras, 439, 123 
\bibitem[P{\'e}nin et al.(2014)]{2014MNRAS.439..143P} P{\'e}nin, A., Lacasa, F., \& Aghanim, N.\ 2014, \mnras, 439, 143 
\bibitem[Planck Collaboration et al.(2014)]{2014A&A...571A..30P} Planck Collaboration, Ade, P.~A.~R., Aghanim, N., et al.\ 2014, \aap, 571, A30  
\bibitem[P{\'e}nin et al.(2012)]{2012A&A...537A.137P} P{\'e}nin, A., Dor{\'e}, O., Lagache, G., \& B{\'e}thermin, M.\ 2012, \aap, 537, A137 
\bibitem[B{\'e}thermin et al.(2011)]{2011A&A...529A...4B} B{\'e}thermin, M., Dole, H., Lagache, G., Le Borgne, D., \& Penin, A.\ 2011, \aap, 529, A4 
\bibitem[Tinker \& Wetzel(2010)]{2010ApJ...719...88T} Tinker, J.~L., \& Wetzel, A.~R.\ 2010, \apj, 719, 88 
\bibitem[Addison et al.(2013)]{2013MNRAS.436.1896A} Addison, G.~E., Dunkley, J., \& Bond, J.~R.\ 2013, \mnras, 436, 1896 
\bibitem[Planck Collaboration et al.(2016)]{2016A&A...596A.109P} Planck Collaboration, Aghanim, N., Ashdown, M., et al.\ 2016, \aap, 596, A109 
\bibitem[Wu \& Dor{\'e}(2017)]{2017MNRAS.466.4651W} Wu, H.-Y., \& Dor{\'e}, O.\ 2017, \mnras, 466, 4651 
\bibitem[Wu et al.(2018)]{2018MNRAS.tmp...57W} Wu, H.-Y., Dor{\'e}, O., Teyssier, R., \& Serra, P.\ 2018, \mnras
\bibitem[Pullen et al.(2017)]{2017arXiv170706172P} Pullen, A.~R., Serra, P., Chang, T.-C., Dore, O., \& Ho, S.\ 2017, arXiv:1707.06172 
\bibitem[Schmidt et al.(2015)]{2015MNRAS.446.2696S} Schmidt, S.~J., M{\'e}nard, B., Scranton, R., et al.\ 2015, \mnras, 446, 2696 
\bibitem[Planck Collaboration et al.(2016)]{2016A&A...594A...1P} Planck Collaboration, Adam, R., Ade, P.~A.~R., et al.\ 2016, \aap, 594, A1 
\bibitem[Mittal et al.(2017)]{2017arXiv170806365M} Mittal, A., de Bernardis, F., \& Niemack, M.~D.\ 2017, arXiv:1708.06365 
\bibitem[B{\'e}thermin et al.(2012)]{2012ApJ...757L..23B} B{\'e}thermin, M., Daddi, E., Magdis, G., et al.\ 2012, \apjl, 757, L23 
\bibitem[B{\'e}thermin et al.(2013)]{2013A&A...557A..66B} B{\'e}thermin, M., Wang, L., Dor{\'e}, O., et al.\ 2013, \aap, 557, A66 

\bibitem[Abazajian et al.(2016)]{2016arXiv161002743A} Abazajian, K.~N., Adshead, P., Ahmed, Z., et al.\ 2016, arXiv:1610.02743  
\bibitem[Schaan et al.(2017)]{2017PhRvD..95l3512S} Schaan, E., Krause, E., Eifler, T., et al.\ 2017, \prd, 95, 123512 

\bibitem[Cooray(2004)]{2004NewA....9..173C} Cooray, A.\ 2004, \na, 9, 173 
\bibitem[van Engelen et al.(2014)]{2014ApJ...786...13V} van Engelen, A., Bhattacharya, S., Sehgal, N., et al.\ 2014, \apj, 786, 13 
\bibitem[Osborne et al.(2014)]{2014JCAP...03..024O} Osborne, S.~J., Hanson, D., \& Dor{\'e}, O.\ 2014, \jcap, 3, 024 
\bibitem[Zahn \& Zaldarriaga(2006)]{2006ApJ...653..922Z} Zahn, O., \& Zaldarriaga, M.\ 2006, \apj, 653, 922 
\bibitem[Pourtsidou \& Metcalf(2015)]{2015MNRAS.448.2368P} Pourtsidou, A., \& Metcalf, R.~B.\ 2015, \mnras, 448, 2368 
\bibitem[Lu \& Pen(2008)]{2008MNRAS.388.1819L} Lu, T., \& Pen, U.-L.\ 2008, \mnras, 388, 1819 
\bibitem[Lu et al.(2010)]{2010PhRvD..81l3015L} Lu, T., Pen, U.-L., \& Dor{\'e}, O.\ 2010, \prd, 81, 123015 
\bibitem[Croft et al.(2017)]{2017arXiv170607870C} Croft, R.~A.~C., Romeo, A., \& Metcalf, R.~B.\ 2017, arXiv:1706.07870 
\bibitem[Metcalf et al.(2017)]{2017arXiv170608939M} Metcalf, R.~B., Croft, R.~A.~C., \& Romeo, A.\ 2017, arXiv:1706.08939 
\bibitem[LSST Science Collaboration et al.(2009)]{2009arXiv0912.0201L} LSST Science Collaboration, Abell, P.~A., Allison, J., et al.\ 2009, arXiv:0912.0201 
\bibitem[Planck Collaboration et al.(2016)]{2016A&A...594A..13P} Planck Collaboration, Ade, P.~A.~R., Aghanim, N., et al.\ 2016, \aap, 594, A13 
\bibitem[Millea et al.(2017)]{2017arXiv170806753M} Millea, M., Anderes, E., \& Wandelt, B.~D.\ 2017, arXiv:1708.06753 
\bibitem[Prince et al.(2017)]{2017arXiv170902227P} Prince, H., Moodley, K., Ridl, J., \& Bucher, M.\ 2017, arXiv:1709.02227 
\bibitem[Yang \& Zhang(2011)]{2011MNRAS.415.3485Y} Yang, X., \& Zhang, P.\ 2011, \mnras, 415, 3485 
\bibitem[Schmidt et al.(2012)]{2012ApJ...744L..22S} Schmidt, F., Leauthaud, A., Massey, R., et al.\ 2012, \apjl, 744, L22 
\bibitem[Scranton et al.(2005)]{2005ApJ...633..589S} Scranton, R., M{\'e}nard, B., Richards, G.~T., et al.\ 2005, \apj, 633, 589
\bibitem[Huff \& Graves(2011)]{2011arXiv1111.1070H} Huff, E.~M., \& Graves, G.~J.\ 2011, arXiv:1111.1070

\bibitem[Hall et al.(2013)]{2013PhRvD..87f4026H} Hall, A., Bonvin, C., \& Challinor, A.\ 2013, \prd, 87, 064026 
\bibitem[Kesden et al.(2003)]{2003PhRvD..67l3507K} Kesden, M., Cooray, A., \& Kamionkowski, M.\ 2003, \prd, 67, 123507 
\bibitem[Hanson et al.(2011)]{2011PhRvD..83d3005H} Hanson, D., Challinor, A., Efstathiou, G., \& Bielewicz, P.\ 2011, \prd, 83, 043005 
\bibitem[B{\"o}hm et al.(2016)]{2016PhRvD..94d3519B} B{\"o}hm, V., Schmittfull, M., \& Sherwin, B.~D.\ 2016, \prd, 94, 043519  
\bibitem[Ferraro \& Hill(2017)]{2017arXiv170506751F} Ferraro, S., \& Hill, J.~C.\ 2017, arXiv:1705.06751 
\bibitem[Namikawa et al.(2013)]{2013MNRAS.431..609N} Namikawa, T., Hanson, D., \& Takahashi, R.\ 2013, \mnras, 431, 609 
\bibitem[Smith et al.(2007)]{2007PhRvD..76d3510S} Smith, K.~M., Zahn, O., \& Dor{\'e}, O.\ 2007, \prd, 76, 043510 
\bibitem[Sherwin \& Das(2010)]{2010arXiv1011.4510S} Sherwin, B.~D., \& Das, S.\ 2010, arXiv:1011.4510 
\bibitem[Kovetz et al.(2017)]{2017arXiv170909066K} Kovetz, E.~D., Viero, M.~P., Lidz, A., et al.\ 2017, arXiv:1709.09066 
\bibitem[Furlanetto et al.(2006)]{2006PhR...433..181F} Furlanetto, S.~R., Oh, S.~P., \& Briggs, F.~H.\ 2006, \physrep, 433, 181 
\bibitem[Seo \& Hirata(2016)]{2016MNRAS.456.3142S} Seo, H.-J., \& Hirata, C.~M.\ 2016, \mnras, 456, 3142 
\bibitem[Zhu et al.(2016)]{2016arXiv161007062Z} Zhu, H.-M., Pen, U.-L., Yu, Y., \& Chen, X.\ 2016, arXiv:1610.07062 
\bibitem[Takada \& Hu(2013)]{2013PhRvD..87l3504T} Takada, M., \& Hu, W.\ 2013, \prd, 87, 123504 
\bibitem[Chiang et al.(2014)]{2014JCAP...05..048C} Chiang, C.-T., Wagner, C., Schmidt, F., \& Komatsu, E.\ 2014, \jcap, 5, 048 
\bibitem[Schaan et al.(2014)]{2014PhRvD..90l3523S} Schaan, E., Takada, M., \& Spergel, D.~N.\ 2014, \prd, 90, 123523 
\bibitem[Scoccimarro et al.(1999)]{1999ApJ...527....1S} Scoccimarro, R., Zaldarriaga, M., \& Hui, L.\ 1999, \apj, 527, 1 



\end{thebibliography}
\end{document}